\RequirePackage[2019-10-01]{latexrelease}
\providecommand{\DeclareCommandCopy}[2]{\let#1#2}
\documentclass[a4paper,fleqn,twoside,twocolumn,nofootinbib,showkeys]{revtex4}

\makeatletter
\providecommand{\@raw@classoptionslist}{}
\makeatother
\usepackage[utf8]{inputenc}
\usepackage{cmap}
\usepackage[nocpr]{ujp}
\fontencoding{T1}\selectfont
\usepackage{mathtools,booktabs,array}
\usepackage{refcount}
\usepackage[hidelinks]{hyperref}
\AtBeginDocument{}
\makeatletter
\newcommand{\UJPplainref}[1]{%
  \@ifundefined{r@#1}{\G@refundefinedtrue\textbf{??}}{\getrefnumber{#1}}}
\newcommand{\UJPlinkref}[1]{%
  \@ifundefined{r@#1}{\G@refundefinedtrue\textbf{??}}{\hyperref[#1]{\getrefnumber{#1}}}}
\AtBeginDocument{\DeclareRobustCommand{\ref}{\@ifstar\UJPplainref\UJPlinkref}}
\def\@author@def#1#2{\move@AU\move@AF\move@AUAF\let\@AF@join\@author@join #1\def\@author{{#2}{}}}
\ifdefined\ArticleUA\else
 \def\flagpacs{00}
 \def\udkpacs{UDC}
 \def\udkpacsnum{\@udk}
\fi
\def\@evenfoot{\small\bfseries\thepage\hfil}
\def\@oddfoot{\hfil\small\bfseries\thepage}
\makeatother
\DeclareMathOperator{\Ei}{Ei}

\newcommand{\dd}{\,\mathrm{d}}
\newcommand{\ii}{\mathrm{i}}
\newcommand{\ee}{\mathrm{e}}
\newcommand{\vct}[1]{\bm{#1}}
\newcommand{\I}{\mathcal{I}}
\newcommand{\F}{\mathcal{F}}
\newcommand{\K}{\mathcal{K}}
\newcommand{\A}{\mathcal{A}}
\newcommand{\Y}{\mathcal{Y}}
\newcommand{\E}{\mathcal{E}}
\newcommand{\GeV}{\mathrm{GeV}}
\newcommand{\fm}{\mathrm{fm}}
\newcommand{\mb}{\mathrm{mb}}
\newcommand{\hc}{\hbar c}

\newcommand{\eik}{\rm eik}
\newcommand{\coh}{\rm coh}
\newcommand{\remn}{\rm rem}
\newcommand{\ptc}{\rm pt}
\newcommand{\Carb}{\ensuremath{{}^{12}\mathrm C(d,p)X}}
\allowdisplaybreaks[2]
\newcommand{\AbsUA}{Досліджено, чи можуть віддача та кулонівська взаємодія пояснити надлишок протонів при великих поздовжніх імпульсах у реакції \(^{12}\mathrm C(d,p)X\). Для цього в моделі Глаубера--Ситенка враховано поздовжню передачу імпульсу й поперечний імпульс зареєстрованого протона. При імпульсі пучка 9,1 ГеВ/\(c\) та поздовжньому імпульсі протона 0,325 ГеВ/\(c\) у системі спокою початкового дейтрона інваріантний диференціальний переріз з урахуванням віддачі становить \(2{,}8\)--\(3{,}6\) мб/ГеВ\(^2\) проти \(18{,}5\) мб/ГеВ\(^2\) за опублікованими даними. У цій області кулонівська поправка до суми каналів мала. Натомість поблизу основного максимуму при нульовому куті когерентний переріз зростає на \(4\)--\(12\%\), а сума когерентного й спектаторного внесків у поперечному наближенні змінюється менш ніж на \(0{,}2\%\). Спектри на вуглеці та водні при 3,5 ГеВ/\(c\) зіставлено з вимірюваннями Ситніка за різних поперечних відборів. Розглянуті ефекти не пояснюють надлишку при великих імпульсах. Щоб установити його кваркове походження, потрібні додаткові спостережувані.}
\newcommand{\AbsEN}{We investigate whether recoil and the Coulomb interaction can explain the excess of protons at large longitudinal momenta in \(^{12}\mathrm C(d,p)X\). We calculate the spectrum in the Glauber--Sitenko model, retaining longitudinal momentum transfer and the transverse momentum of the detected proton. At a beam momentum of 9.1 GeV/\(c\) and a longitudinal proton momentum of 0.325 GeV/\(c\) in the rest frame of the incident deuteron, the invariant differential cross section with recoil is \(2.8\)--\(3.6\) mb/GeV\(^2\), against \(18.5\) mb/GeV\(^2\) in the published spectrum. The Coulomb correction to the channel sum is small in this region. Near the main peak at zero proton angle, the coherent cross section rises by \(4\)--\(12\%\), while the sum of coherent breakup and neutron removal changes by less than \(0.2\%\) in the transverse approximation. At 3.5 GeV/\(c\) we compare the calculated carbon and hydrogen spectra with Sitnik's measurements for several transverse cuts. The effects considered here do not account for the high-momentum excess. Further observables are needed to establish whether it has a quark origin.}
\begin{document}
\title[High-momentum protons from deuteron breakup]{Forward deuteron breakup on carbon and hydrogen: Coulomb effects and the high-momentum excess}
\ifdefined\ReviewCopy
\author{ }\affiliation{ }\autorcol{ }
\else
\author{Ya.~D.~KRIVENKO-EMETOV}
\affiliation{National Technical University of Ukraine “Igor Sikorsky Kyiv Polytechnic Institute”, 37 Beresteiskyi Ave., Kyiv 03056, Ukraine; e-mail: y.kryvenko-emetov@kpi.ua}
\affiliation{Institute for Nuclear Research, National Academy of Sciences of Ukraine, 47 Prospekt Nauky, Kyiv 03028, Ukraine}
\author{B.~I.~SYDORENKO}
\affiliation{Former member of the Institute for Nuclear Research, National Academy of Sciences of Ukraine, 47 Prospekt Nauky, Kyiv 03028, Ukraine; e-mail: boris.sidorenko@meta.ua}
\autorcol{Ya.~D.~Krivenko-Emetov, B.~I.~Sydorenko}
\fi
\begin{abstract}
\AbsEN
\end{abstract}
\udk{539.125.17+539.126.17}
\maketitle
\section{Introduction}
The proton spectrum from forward deuteron breakup on carbon has two regions of interest. Its main peak occurs at a laboratory proton momentum close to half the beam momentum. At large longitudinal momenta in the rest frame of the incident deuteron, the cross section falls, but the measured yield exceeds the nucleon-based prediction \cite{Ableev1983,Kobushkin2008}. This excess has been associated with a six-quark admixture or with quark exchange \cite{Ableev1983,Kobushkin1998}. The purpose of the present work is to determine how much of it can be explained by the breakup kinematics and the Coulomb interaction within a nucleon model.

We describe single and double scattering in the Glauber--Sitenko approximation \cite{Glauber1959,Akhiezer1957,Sitenko1991}. Glauber calculations of breakup spectra at finite proton angles date back to Ref.~\cite{Azhgirey1978}. An early diffraction treatment of longitudinal transfer was given by Tartakovsky, Fursayev, and Sidorenko for triton breakup by a proton \cite{Tartakovsky2005}. The effects of longitudinal transfer and transverse relative momentum on deuteron breakup on hydrogen were examined in Ref.~\cite{hydrogen}. Here we retain both the laboratory longitudinal transfer \(Q_z\) and the detected proton's transverse momentum \(p_t\) when including the Coulomb interaction. Fixing \(p_t\) leaves the internal momentum of the \(pn\) pair dependent on the recoil.

In \(A(d,p)X\) only the proton is observed; the inclusive spectrum sums over the unobserved final states \(X\). In the carbon calculation we distinguish coherent breakup, in which the target remains in its ground state, from the yield of spectator protons after nonelastic scattering of the neutron. The Coulomb effect in this reaction was considered in Refs.~\cite{Kobushkin2008,Krivenko2021,preprint}. We calculate the correction to each contribution with the unexpanded phase and compare it with the expansion in the Sommerfeld parameter. With real optical profiles the linear correction to the cross section vanishes, so the second order is needed.

At a beam momentum of 9.1 GeV/\(c\), we examine the high-momentum part of the forward carbon spectrum using a longitudinal transfer determined by recoil kinematics. Sitnik's measurements with the Strela spectrometer at 3.5 GeV/\(c\) \cite{Sitnik2019} provide a separate test of the transverse dependence near the main peak on carbon and hydrogen.

\section{Reaction model}\label{sec:model}
\subsection{Kinematics and transition density}
\begin{figure}[t]
 \centering\includegraphics[width=\columnwidth]{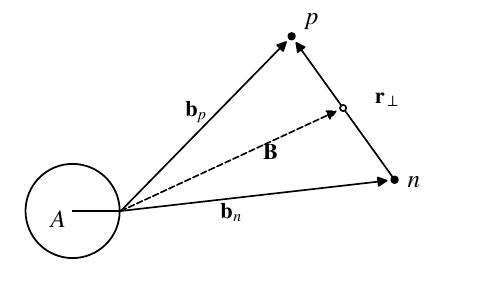}
 \caption{Impact parameters in the plane normal to the beam. The vectors \(\vct b_p\) and \(\vct b_n\) join the target centre to the proton and neutron, respectively; \(\vct r_\perp\) points from neutron to proton.}
 \label{fig:geom}
\end{figure}

Let the beam define the \(z\) axis. A deuteron of mass \(M_d\) and laboratory momentum \(p_d\) has speed \(v_d=p_d/\sqrt{p_d^2+M_d^2}\) and Lorentz factor \(\gamma_d=(1-v_d^2)^{-1/2}\); we use \(c=1\) unless units are shown explicitly. For a detected proton of momentum \((\vct p_\perp,p_z^{\rm lab})\) and energy \(E_p^{\rm lab}\), the longitudinal momentum in the initial-deuteron rest frame (asterisk) is
\begin{equation}
 k_z=p_z^*=\gamma_d(p_z^{\rm lab}-v_d E_p^{\rm lab}).
 \label{eq:boost}
\end{equation}
In the transverse approximation the momentum transferred to the pair is \((\vct Q,0)\), with \(\vct Q\) in the plane perpendicular to the beam. For equal nucleon masses the relative momentum is \(\vct k=(\vct p_\perp-\vct Q/2,k_z)\). At zero angle we set \(Q_z=p_\perp=0\), so the internal momentum \(\vct k_\perp=-\vct Q/2\) still varies in the coherent integral. The label \(k_z\) refers to the incident-deuteron frame: in this approximation we neglect longitudinal momentum and energy transfer in forming the boost of the pair. Section~\ref{sec:finite} considers nonzero transfers. The corresponding diffraction kinematics for hydrogen was discussed in Ref.~\cite{hydrogen}.
The relative coordinate \(\vct r=(\vct r_\perp,r_z)\) points from the neutron to the proton. If the deuteron centre passes the target at impact parameter \(\vct B\) (Fig.~\ref{fig:geom}), the nucleon impact parameters are \(\vct b_p=\vct B+\vct r_\perp/2\) and \(\vct b_n=\vct B-\vct r_\perp/2\). The incident deuteron is represented by a normalized scalar \(S\) wave,
\begin{equation}
 \begin{aligned}
 \psi_S(r)&=\frac{1}{2\sqrt{\pi}N_r}\sum_j A_j\ee^{-\lambda_j r^2},\\
 \int\dd^3r\,|\psi_S(r)|^2&=1 .
 \end{aligned}
 \label{eq:wave}
\end{equation}
The real coefficients \(A_j\) and positive widths \(\lambda_j\) of the twelve-Gaussian K2 wave function are given in Appendix~\ref{app:gauss}, and \(N_r\) fixes its norm. In the coherent amplitude the outgoing plane wave is made orthogonal to \(\psi_S\) by projecting out the bound state. Its conjugate is \(\phi_{\vct k}^*=\ee^{\ii\vct k\cdot\vct r}-W(\vct k)\psi_S(r)\), where \(W\) is the unnormalized Fourier transform of \(\psi_S\). Denote the transform of \(|\psi_S|^2\) in the same convention by \(n_S\). The transition form factor is then
\begin{equation}
 \begin{split}
 G(\vct q,\vct k)&=\int\dd^3r\,\phi_{\vct k}^*
                      \ee^{\ii\vct q\cdot\vct r}\psi_S(r)\\
  &=W(\vct k+\vct q)-W(\vct k)n_S(\vct q),\\[-2pt]
 G(0,\vct k)&=0.
 \end{split}
 \label{eq:ff}
\end{equation}
Since \(n_S(0)=1\), the form factor vanishes at zero transfer. The subtraction eliminates the overlap with the bound deuteron; it does not describe the \(pn\) final-state interaction.
\subsection{Optical profiles and the Coulomb phase}
For nucleon $N=p,n$ the strong transmission factor is $S_N(b)=1-\Gamma_N(b)$, with
\begin{equation}
 \begin{gathered}
 \Gamma_N(b)=C_N\ee^{-d_N b^2},\qquad d_N=\frac{1}{2\beta_N^2},\\
 C_N=\frac{(1-\ii\rho_N)\sigma_N}{4\pi\beta_N^2}.
 \end{gathered}
 \label{eq:profiles}
\end{equation}
For a nucleon of momentum \(p_N\), the profile is specified by the total nucleon--carbon cross section \(\sigma_N\), the transverse width \(\beta_N\), and the ratio \(\rho_N\) of the real and imaginary parts of the forward strong amplitude. With \(g_N=\beta_N^2/2\) the amplitude in this convention is \(f_N(Q)=p_N(\ii+\rho_N)\sigma_N\ee^{-g_N Q^2}/(4\pi)\). We neglect spin dependence and keep the internal coordinates fixed during passage through the target.

The target charge is represented by a harmonic-oscillator form factor normalized to unity at zero transfer \cite{Kobushkin2008}. After subtracting the common constant from the screened Coulomb phase, the transverse phase is
\begin{align}
 F_{\rm ch}(Q)&=(1-\kappa a_0^2Q^2/2)\ee^{-a_0^2Q^2/4}, \label{eq:charge}\\
 t(b)&=c_C+2\ln(b/b_0)+E_1(b^2/a_0^2)
                   +2\kappa\ee^{-b^2/a_0^2}, \label{eq:phase}\\
 \eta&=\frac{Z\alpha_{\rm em}}{v_d},\qquad
 \kappa=\frac{Z-2}{3Z}. \nonumber
\end{align}
Here \(Z=6\) for carbon, \(\alpha_{\rm em}\) is the fine-structure constant, \(a_0\) is the charge length, \(b_0\) is an arbitrary reference length, and \(c_C\) is a real constant. The exponential integral is \(E_1(x)=\int_x^\infty \ee^{-s}\dd s/s\) for \(x>0\). The dimensionless oscillator coefficient \(\kappa\) enters the rms charge radius as \(r_{\rm ch}=a_0\sqrt{3/2+3\kappa}\). At the origin the logarithmic singularities of the first two \(b\)-dependent terms cancel, giving \(t(0)=c_C+2\ln(a_0/b_0)-\gamma_E+2\kappa\), where \(\gamma_E\) is Euler's constant. The phase tends to its point-charge form at large \(b\). In both breakup channels the Coulomb phase multiplies the proton transmission factor.

Define the coherent amplitude $\F_\eta=F_\eta/p_d$ by
\begin{align}
 \F_\eta &=\frac{\ii}{2\pi}\int\dd^2B\,
   \ee^{\ii\vct Q\cdot\vct B}
   \int\dd^3r\,\phi_{\vct k}^*\psi_S(r) \notag\\
 &\quad\times
 [1-S_p(b_p)S_n(b_n)\ee^{\ii\eta t(b_p)}].
 \label{eq:amp}
\end{align}
Simultaneous inversion of \(\vct B\) and \(\vct r\) leaves the scalar reaction operator and incident state invariant; it gives the positive signs in both Fourier exponents. At \(\eta=0\) the expansion of the product of transmission factors gives the proton and neutron single-scattering terms and their double-scattering term with signs \(+,+,-\). The identity term in Eq.~\eqref{eq:amp} has zero matrix element between the bound and projected states. A constant shift \(t\mapsto t+\tau\) therefore multiplies \(\F_\eta\) by \(\ee^{\ii\eta\tau}\) and leaves the spectrum invariant.
\section{Evaluation of the coherent amplitude}\label{sec:method}
\subsection{Single scattering and rescattering}
Use the transverse Fourier transform \(\widetilde f(\vct\ell)=\int\dd^2b\,\ee^{\ii\vct\ell\cdot\vct b}f(\vct b)\) with inverse measure \(\dd^2\ell/(2\pi)^2\). The following reduction holds for the unexpanded phase and for the individual orders:
\begin{align}
 G_\pm&=G((\pm\vct Q/2,0),\vct k), \notag\\
 \Phi(\vct\ell)&=\ee^{-g_n|\vct Q-\vct\ell|^2}
          G((\vct Q/2-\vct\ell,0),\vct k), \label{eq:phi}\\
 D_\eta(b)&=\ee^{\ii\eta t(b)}-1,\qquad
 U_\eta(b)=S_p(b)D_\eta(b). \notag
\end{align}
\begin{align}
 \F_0={}&\ii C_p\beta_p^2\ee^{-g_p Q^2}G_-
           +\ii C_n\beta_n^2\ee^{-g_n Q^2}G_+
           \notag\\
 &-\frac{\ii C_p C_n\beta_p^2\beta_n^2}{2\pi}
       \int\dd^2\ell\,\ee^{-g_p\ell^2}\Phi(\vct\ell), \label{eq:fn}\\
 \F_\eta={}&\F_0-\frac{\ii}{2\pi}\widetilde U_\eta(Q)G_-
     +\frac{\ii C_n\beta_n^2}{4\pi^2}
         \int\dd^2\ell\,\widetilde U_\eta(\ell)\Phi(\vct\ell).
 \label{eq:feta}
\end{align}
Expanding \(-D_\eta S_p(1-\Gamma_n)=-U_\eta+U_\eta\Gamma_n\) in Eq.~\eqref{eq:amp} fixes the sign of the last term. Both Coulomb terms thus contain the proton profile. The strong convolution in Eq.~\eqref{eq:fn} involves only Gaussians; the widths and linear vectors for both convolutions are given in Appendix~\ref{app:gauss}.

\subsection{Analytic angular integration and radial quadrature}
Each term of \(\Phi\) is proportional to \(w(\vct\ell)=\exp(-a\ell^2+\vct L\cdot\vct\ell)\), with \(a>0\) and real \(\vct L\). In the last term of Eq.~\eqref{eq:feta}, write the transform as an integral over \(\vct b\) and first integrate over \(\vct\ell\):
\begin{align}
 \int\dd^2\ell\,w(\vct\ell)\ee^{\ii\vct\ell\cdot\vct b}
  &=\frac{\pi}{a}\ee^{(L^2-b^2)/(4a)}
                 \ee^{\ii\vct L\cdot\vct b/(2a)},\notag\\
 L&=|\vct L| .\label{eq:complete}
\end{align}
Completing the square gives this two-dimensional transform directly. For radial \(f(b)\), the azimuthal integral of the remaining Fourier phase is \(2\pi J_0(Lb/(2a))\). Setting \(x=b/(2\sqrt a)\) yields
\begin{align}
 \mathcal M_f(a,L)&=\int\dd^2\ell\,w(\vct\ell)\widetilde f(\vct\ell)
   \notag\\
 &=8\pi^2\ee^{L^2/(4a)}
   \int_0^\infty x\dd x\,\ee^{-x^2}\notag\\[-2pt]
 &\quad\times J_0(Lx/\sqrt a) f(2\sqrt a\,x).
 \label{eq:moment}
\end{align}
Here \(J_0\) is the Bessel function of order zero. Even without screening, Eq.~\eqref{eq:moment} pairs the Fourier transform of \(U_\eta\) with a Gaussian. The same formula applies to \(tS_p\) and \(t^2S_p\), because the Gaussian bounds their growth at large \(b\). The result retains the dependence on \(k_z\) and \(\beta_N\).

In momentum space the azimuth enters the argument of the exponential integral through \(\vct Q\cdot\vct k_\perp\). For real arguments \(X_1X_2\ne0\) we can write the difference as
\begin{align}
 \Ei(X_1)-\Ei(X_2)&=\ln|X_1/X_2| \notag\\
 &\quad+\int_0^1\frac{\ee^{uX_1}-\ee^{uX_2}}{u}\dd u .
 \label{eq:eitwo}
\end{align}
The numerator of the integral vanishes at \(u=0\). At a common zero of the two arguments, the logarithmic difference has a finite limit if their ratio tends to a finite nonzero value. At distinct zeros the logarithms remain; each is locally integrable in the transverse plane. Equation~\eqref{eq:moment} carries out the angular integration before the radial quadrature, where only the Gaussian-weighted functions are needed. The analytic first-order moments are derived in Appendix~\ref{app:moments}.

The bare proton term needs a separate transform because \(D_\eta\) persists at large \(b\). Split the phase into the point-charge part \(t_{\ptc}(b)=c_C+2\ln(b/b_0)\) and the difference \(\delta t(b)=t(b)-t_{\ptc}(b)\). For \(Q>0\) the point-charge transform is \cite{DLMF}
\begin{align}
 \widetilde D_{\eta,\ptc}(Q)&=
 \frac{-4\pi\ii\eta}{Q^2}
 \frac{\Gamma(1+\ii\eta)}{\Gamma(1-\ii\eta)}
 \ee^{\ii\eta L_C(Q)}, \notag\\
 L_C(Q)&=c_C+2\ln[2/(Qb_0)].
 \label{eq:point}
\end{align}
Here \(\Gamma(z)\) denotes the Euler gamma function. Equation~\eqref{eq:point} uses the distributional continuation of the Mellin integral of \(J_0\). The finite-size correction has integrand \(\ee^{\ii\eta t_{\ptc}}(\ee^{\ii\eta\delta t}-1)\), and the nuclear profile gives a separate term \(-C_p\ee^{-d_p b^2}D_\eta\). Both use the Hankel transform \(2\pi\int_0^\infty b\dd b\,J_0(Qb)f(b)\) and fall rapidly at large \(b\). The finite-size difference is bounded as \(b\to0\) for real \(\eta\); its first two Taylor coefficients remain integrable. At zero transfer the distributional term drops out because \(G_-(0)=0\). At \(p_\perp=0\) the radial \(S\) wave also gives \(G_-=O(Q^2)\), by expansion of Eq.~\eqref{eq:ff} about \(\vct k_\perp=-\vct Q/2=0\). We evaluate \(G_-/Q^2\) before multiplying by the phase and integrate with measure \(2\pi Q\dd Q\) down to \(Q=0\).

At nonzero \(\vct p_\perp\) the transverse gradient of \(W\) need not vanish, and \(G_-\) can begin at order \(Q\). Squaring the Coulomb amplitude then makes the small-\(Q\) region sensitive to the longitudinal transfer and excitation energy. The finite-transfer treatment of the external pole is given in Sec.~\ref{sec:finite}; it is separate from the Fourier subtraction in the internal convolution.
\section{Proton spectrum and expansion in $\eta$}\label{sec:spec}
\subsection{Coherent breakup and neutron removal}
The inclusive measurement fixes the proton momentum and sums over the rest of the final state. For a channel matrix \(S=1+\ii T\), unitarity gives \(2\,\mathrm{Im}\,T_{ii}=\sum_f\int\dd\Phi_f\,|T_{fi}|^2\), where \(\dd\Phi_f\) is the phase-space measure and the sum includes the elastic channel. The ordinary optical theorem gives the total cross section; subtracting the elastic term gives the nonelastic cross section. To obtain a proton spectrum one must retain the projection onto the observed momentum; in the generalized optical theorem this is expressed by a missing-mass discontinuity \cite{Mueller1970}. In the spectator approximation used here, the sum over neutron final states enters through the local loss probability:
\begin{align}
 P_n(b) &= 1-|S_n(b)|^2
        =2\,\mathrm{Re}\,\Gamma_n(b)-|\Gamma_n(b)|^2, \notag\\
 \int\dd^2b\,P_n(b)&=\sigma_n-
       \frac{\sigma_n^2(1+\rho_n^2)}{16\pi\beta_n^2}.
 \label{eq:loss}
\end{align}
Here \(b=|\vct b_n|\), and \(P_n\) is the probability that the neutron leaves the elastic channel at that impact parameter. Integrating the two Gaussians gives the nonelastic neutron--carbon cross section in this optical model. The differential yield of the surviving proton also depends on its momentum and transmission amplitude.

In the frozen spectator approximation, for a given neutron impact parameter \(\vct b_n\) and \(\vct p_\perp=0\), the amplitude for the surviving proton is
\begin{align}
 \A_\eta(b,k_z)&=\int\dd^3r\,\ee^{-\ii k_zr_z}\psi_S(r)
       S_p(b')\ee^{\ii\eta t(b')}, \label{eq:remamp}\\
 b'&=|\vct b_n+\vct r_\perp|. \notag
\end{align}
The coherent and neutron-removal spectra are
\begin{align}
 \I_{\coh} &= \K \int\dd^2Q\,|\F_\eta|^2,\notag\\
 \I_{\remn} &=\K \int\dd^2b\,P_n(b)|\A_\eta|^2,\notag\\
 \I_{\rm sum}&=\I_{\coh}+\I_{\remn}.
 \label{eq:spectrum}
\end{align}
The two spectra add without interference because the coherent contribution leaves the target in its ground state and neutron removal leads to nonelastic target states. Their sum in Eq.~\eqref{eq:spectrum} is the part of \Carb\ described by this model. At zero angle the invariant spectrum is \(\I=E_p^{\rm lab}\dd^3\sigma/\dd^3p^{\rm lab}=E_p^*\dd^3\sigma/\dd^3p^*\). With lengths in fm, momenta in the integrals in \(\fm^{-1}\), and \(E_p^*=\sqrt{m_p^2+(k_z^{\GeV})^2}\) in GeV, the coefficient in Eq.~\eqref{eq:spectrum} is
\begin{equation}
 \K=\frac{10 E_p^*}{(2\pi)^3(\hc)^3},
 \qquad \hc=0.19733\ \GeV\,\fm .
 \label{eq:units}
\end{equation}
Here \(m_p\) is the proton mass and \(k_z^{\GeV}\) is \(k_z\) in GeV/\(c\). The factor \(10\) converts \(\fm^2\) to mb, \((\hc)^{-3}\) converts the momentum density, and \((2\pi)^{-3}\) follows from the normalization of the continuum. Thus \(\I\) is expressed in \(\mb\,\GeV^{-2}\) at \(c=1\).

After integrating over \(r_z\) for each Gaussian in Eq.~\eqref{eq:remamp}, change variables to \(\vct b'=\vct b_n+\vct r_\perp\). Integration over the azimuth gives \(2\pi I_0(2\lambda_jbb')\), where \(I_0\) is the modified Bessel function of order zero. Thus
\begin{align}
 \A_\eta(b,k_z)&=\sum_j\frac{A_j}{2\sqrt{\pi}N_r}
              \sqrt{\frac{\pi}{\lambda_j}}\,
              \ee^{-k_z^2/(4\lambda_j)} R_j(b),\notag\\
 R_j(b)&=2\pi\int_0^\infty b'\dd b'\,
        \ee^{-\lambda_j(b'^2+b^2)}\notag\\[-2pt]
 &\quad\times I_0(2\lambda_jbb') S_p(b')\ee^{\ii\eta t(b')}.
 \label{eq:radrem}
\end{align}
For numerical evaluation set \(y=2\lambda_jbb'\) and factor the weight as \(\ee^{-\lambda_j(b'^2+b^2)}I_0(y)=\ee^{-\lambda_j(b'-b)^2}[\ee^{-y}I_0(y)]\). The scaled Bessel function avoids large intermediate values. At \(\eta=0\) the two terms of \(R_j\) reduce to Gaussian moments:
\begin{equation}
 R_j^{[0]}(b)=\frac{\pi}{\lambda_j}
  -\frac{\pi C_p}{\lambda_j+d_p}
       \exp\!\left[-\frac{\lambda_jd_p b^2}{\lambda_j+d_p}\right].
 \label{eq:remclosure}
\end{equation}
Equation~\eqref{eq:remclosure} checks the inner integral without quadrature. The outer integral in Eq.~\eqref{eq:spectrum} extends over the neutron radius \(b\).

\subsection{Expansion of the cross section through second order}
For either channel write \(\Y_\eta=\Y_0+\Y_1+\Y_2+\cdots\), with the powers \(\eta\) and \(\eta^2\) included in \(\Y_1\) and \(\Y_2\). The two terms follow by inserting \(\ii\eta t\) and \(-\eta^2t^2/2\) in Eqs.~\eqref{eq:feta} and \eqref{eq:radrem}. For example, at \(Q>0\) the quadratic point-charge term in \(\widetilde D_{\eta,\ptc}\) is \(4\pi\eta^2[L_C(Q)-2\gamma_E]/Q^2\). Its finite-size companion has integrand \(-\eta^2\delta t(t_{\ptc}+\delta t/2)\). The proton profile and the double-scattering term weight \(-\eta^2t^2/2\) in their respective integrals. The squared amplitude through order \(\eta^2\) is
\begin{align}
 |\Y|^2_{[0]} &=|\Y_0|^2,\notag\\
 |\Y|^2_{[1]} &=|\Y|^2_{[0]}+2\,\mathrm{Re}(\Y_0^*\Y_1),\notag\\
 |\Y|^2_{[2]} &=|\Y|^2_{[1]}+|\Y_1|^2+2\,\mathrm{Re}(\Y_0^*\Y_2).
 \label{eq:orders}
\end{align}
Inserting Eq.~\eqref{eq:orders} into Eq.~\eqref{eq:spectrum} gives \(\I^{[m]}\); for \(\I^{\eik}\) we retain the full phase in \(\Y_\eta\). Squaring \(\Y_0+\Y_1\) includes just one of the two terms of order \(\eta^2\). To see why both are needed, shift the phase by a constant: for \(t\mapsto t+\tau\), \(\Y_1\mapsto\Y_1+\ii\eta\tau\Y_0\) and \(\Y_2\mapsto\Y_2+\ii\eta\tau\Y_1-(\eta\tau)^2\Y_0/2\). The changes cancel at each order in Eq.~\eqref{eq:orders}. In particular, a constant \(t\) leaves the cross section unchanged.

For real profiles (\(\rho_p=\rho_n=0\)), conjugation in Eq.~\eqref{eq:amp} followed by \(\vct B\mapsto-\vct B\) and \(\vct r\mapsto-\vct r\) gives \(\F_{-\eta}=-\F_\eta^*\). In the spectator amplitude~\eqref{eq:remamp}, evenness in \(r_z\) gives \(\A_{-\eta}=\A_\eta^*\). Each spectrum is then even in \(\eta\). A small real part of the strong amplitude makes the linear interference small without constraining the quadratic term in the same way.
\section{Numerical results}\label{sec:num}
\subsection{Carbon parameters and numerical convergence}
At the nominal beam momentum \(p_d=9.1\) GeV/\(c\) of the published zero-angle spectrum \cite{Kobushkin2008}, the masses \(M_d=1.876\) and \(m_p=0.938\) GeV/\(c^2\) give \(\eta\simeq0.0447\). We use the same normalized K2 \(S\) wave for both optical sets in Table~\ref{tab:inputs}. In set A the charge length follows from the \(^{12}\)C charge radius of \(2.47\) fm \cite{Angeli}; set B takes the optical and charge lengths from Ref.~\cite{Kobushkin2008}, where the momentum-space slope corresponds to \(\beta^2=69.3\) (GeV/\(c\))\(^{-2}\) in Eq.~\eqref{eq:profiles}. In each set the proton and neutron profiles are taken equal. We use the calculated normalization without fitting the breakup spectrum. The phase reference \(c_C=0\), \(b_0=1\) fm has no effect on the yields.
\begin{table}[t]
 \setlength{\belowcaptionskip}{1.5pt}
 \caption{Optical and charge parameters for carbon. The proton and neutron entries are equal within each set.}
 \label{tab:inputs}
 \centering
 \begin{tabular}{@{}lrr@{}}
 \toprule
 Parameter & A & B\\\midrule
 $\sigma_N$ (mb) & 350 & 340\\
 $\beta_N$ (fm) & 2.19 & 1.64\\
 $\rho_N$ & $10^{-4}$ & 0\\
 $a_0$ (fm) & 1.68 & 1.60\\
 $\int P_n\,\dd^2b$ (mb) & 299 & 255\\
 \bottomrule
 \end{tabular}
\end{table}

The numerical integrations are over the radial transforms in Eqs.~\eqref{eq:moment} and \eqref{eq:point}, the coherent transfer \(Q\), and the two radii in the neutron-removal term. We use Gauss--Legendre quadrature with 160 nodes for the Coulomb transforms and the inner integral of Eq.~\eqref{eq:radrem}. The 112 outer nodes in that equation span \(0<b<9\beta_n\). The Gaussian variable \(x\) runs to 9 in Eq.~\eqref{eq:moment} and in the localized proton transforms; in Eq.~\eqref{eq:radrem} the inner endpoint is \(b'=b+10/\sqrt{\lambda_j}\). The coherent integral uses 36 nodes in each \(Q\) interval with endpoints
\begin{equation*}
 (0,\ 0.001,\ 0.005,\ 0.02,\ 0.08,\ 0.20,\ 0.45,\ 0.60,\ 0.80)
\end{equation*}
in GeV/$c$. The subdivisions resolve the soft end of the integral; the first interval includes the origin as an endpoint. The $Q^2$ factor in $G_-$ makes the zero-angle spectrum integrable there.

For eleven values of \(k_z\) between 0 and 0.50 GeV/\(c\), increasing the two radial node counts to 256 and 176, using 54 nodes per \(Q\) interval, and taking \(Q_{\max}=1\) GeV/\(c\) changed the yields by less than \(10^{-7}\) in relative terms for either set. The first-order analytic moments in Appendix~\ref{app:moments} agree with Eq.~\eqref{eq:moment}; direct integration of the exponential integral at positive argument together with its logarithmic term checks the moment weighted by the nuclear profile on both targets. Symmetric differences of the unexpanded complex amplitudes at \(\eta=\pm10^{-5}\) reproduce the first- and second-order coefficients to within a few parts in \(10^9\). With \(c_C\mapsto c_C+1.7\), all calculated yields, including the truncated ones, agree to within \(10^{-8}\). For the removal integrals we also use the analytic limits in Eqs.~\eqref{eq:remclosure} and \eqref{eq:loss}.
\begin{figure*}[t]
 \centering\includegraphics[width=\textwidth]{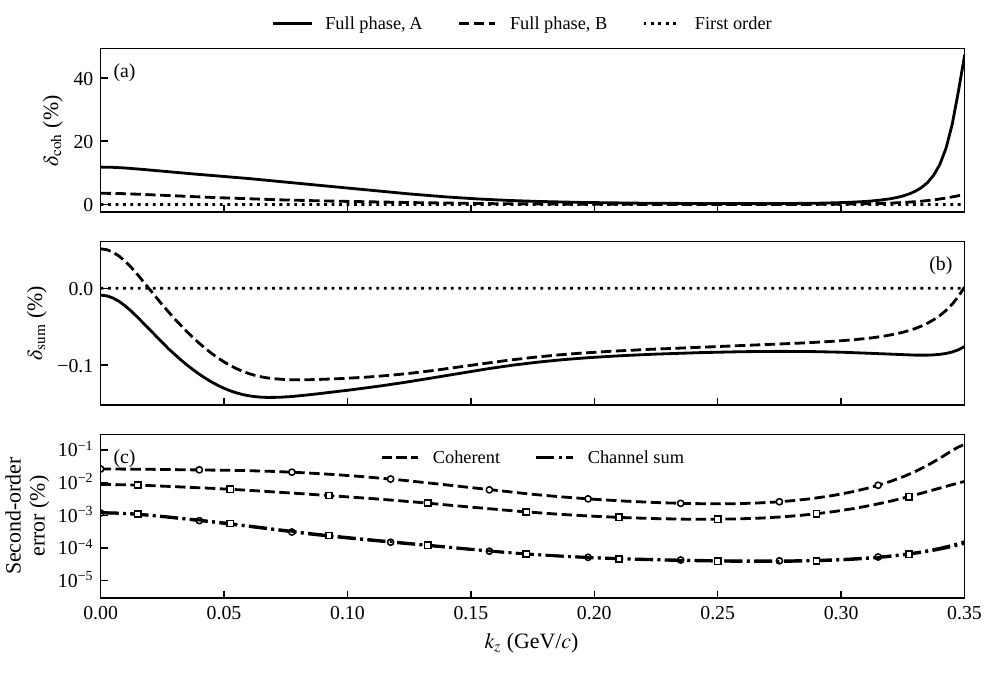}
 \caption{Comparison of the Sommerfeld expansion with the full transverse phase. (a,b) Fractional changes $\delta_x=100(\I_x/\I_x^{[0]}-1)$ for the coherent contribution and the sum of the two channels; the full phase is shown by solid (A) and dashed (B) lines, the strict first order by the dotted line. (c) Absolute relative difference between second order and the full phase for the coherent contribution and for the sum. In (c), circles mark A and squares mark B.}
 \label{fig:orders}
\end{figure*}
\begin{table*}[t]
 \setlength{\belowcaptionskip}{1.5pt}
 \caption{Invariant yields at \(k_z=0\) and Coulomb changes from the strong result \(\I^{[0]}\), in mb/GeV\(^2\) at \(c=1\). The second-order column includes the linear term. The sums are rounded after addition.}
 \label{tab:changes}
 \centering
 \begin{tabular}{@{}clrrrr@{}}
 \toprule
 Set & Contribution & $\I^{[0]}$
 & $\I^{[1]}-\I^{[0]}$ & $\I^{[2]}-\I^{[0]}$ & $\I^{\eik}-\I^{[0]}$\\\midrule
 A & Coherent       & $1.01\times10^4$ & $+0.31$ & $+1.19\times10^3$ & $+1.20\times10^3$\\
   & Neutron removal& $3.96\times10^5$ & $+0.69$ & $-1.23\times10^3$ & $-1.23\times10^3$\\
   & Sum            & $4.06\times10^5$ & $+1.0$ & $-40$ & $-35$\\[1pt]
 B & Coherent       & $3.23\times10^4$ & 0 & $+1.15\times10^3$ & $+1.15\times10^3$\\
   & Neutron removal& $3.22\times10^5$ & 0 & $-970$ & $-969$\\
   & Sum            & $3.54\times10^5$ & 0 & $+179$ & $+183$\\
 \bottomrule
 \end{tabular}
\end{table*}
\subsection{Coulomb effects in the two channels}
At the centre of the spectrum (Table~\ref{tab:changes}), the linear Coulomb interference in the coherent yield for set A is \(0.31\) mb/GeV\(^2\) on a strong yield of about \(10^4\) mb/GeV\(^2\). Nevertheless, the unexpanded phase raises the coherent yield by about \(12\%\) for A and \(3.6\%\) for B. At the same momentum the neutron-removal yield falls by about \(0.3\%\) in both sets. Because neutron removal dominates, the two shifts are comparable in absolute value and have opposite signs. The sum changes by only \(-0.009\%\) for A and \(+0.05\%\) for B. These cancellations depend on the reaction parameters.

At \(k_z=0\) in set A the coherent contribution from \(|\F_1|^2\) is about \(1.47\times10^3\) mb/GeV\(^2\). The interference with \(\F_2\) supplies another contribution of about \(-2.75\times10^2\) mb/GeV\(^2\). Together they give a second-order yield within \(0.03\%\) of the unexpanded value. Omitting the interference with \(\F_2\) would overestimate the Coulomb increase by roughly one quarter for the chosen phase reference.

Figure~\ref{fig:orders} shows how these changes vary with momentum. At the 44 points used for comparison with experiment, \(\I^{[2]}\) and \(\I^{\eik}\) differ by less than \(0.1\%\) of the coherent yield and \(0.002\%\) of the sum in either set. The unexpanded phase changes the sum relative to \(\I^{[0]}\) by less than \(0.2\%\) at those points. These percentages refer to the yields themselves. Near the coherent minimum for set A, a sizeable relative correction can correspond to a very small absolute change.

In set B the profiles are real, so the linear interference vanishes. If we change only \(\rho_p=\rho_n\) in set A to \(0.10\), the linear coherent term at \(k_z=0\) grows to about \(3.1\times10^2\) mb/GeV\(^2\) and the unexpanded phase changes the sum there by about \(0.24\%\). The small Coulomb effect on the sum therefore pertains to the inputs and forward geometry considered here.
\subsection{Comparison with the published carbon spectrum}
\begin{figure*}[t]
 \centering\includegraphics[width=\textwidth]{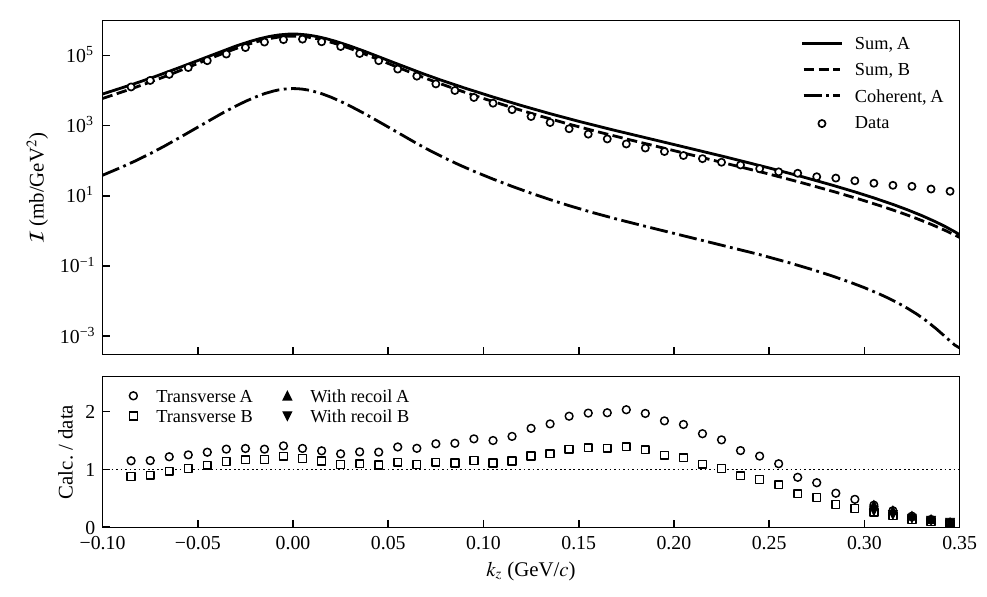}
 \caption{Forward proton spectrum on carbon. The upper panel shows data from Fig.~3 of Ref.~\cite{Kobushkin2008}, the two channel sums in the transverse calculation (A and B), and the coherent contribution alone for A. The lower panel gives the calculation divided by the data: circles and squares are the transverse results, and upward and downward triangles include on-shell recoil for A and B, respectively.}
 \label{fig:data}
\end{figure*}

Figure~\ref{fig:data} compares the calculated spectra with 44 points read from the experimental markers in Fig.~3 of Ref.~\cite{Kobushkin2008}. That figure plots the proton momentum in the incident-deuteron rest frame at \(0^\circ\), which corresponds to \(k_z\) here. We use its linear momentum scale and logarithmic cross-section scale without renormalizing the ordinates. The points are from the measurements cited in that figure, including Ableev \textit{et al.} and Zaporozhets \textit{et al.}~\cite{Ableev1983,Zaporozhets1986}. The calculated curves are for the nominal beam momentum and zero proton angle; the original acceptance and covariance are unavailable for a fit.
\begin{table}[t]
 \setlength{\belowcaptionskip}{1.5pt}
 \caption{Selected points in the transverse approximation. Momenta are in GeV/$c$ and yields in mb/GeV$^2$. Columns A and B give the sums with the full phase.}
 \label{tab:data}
 \centering
 \small
 \setlength{\tabcolsep}{3pt}
 \begin{tabular}{@{}rrrr@{}}
 \toprule
 $k_z$ & Data & A & B\\\midrule
 0.005 & $2.90\times10^5$ & $3.95\times10^5$ & $3.45\times10^5$\\
 0.105 & $4.34\times10^3$ & $6.50\times10^3$ & $4.82\times10^3$\\
 0.205 & $1.40\times10^2$ & $2.49\times10^2$ & $1.68\times10^2$\\
 0.325 & 18.5 & 3.5 & 2.5\\
 \bottomrule
 \end{tabular}
\end{table}
Including neutron removal brings the calculated yield to the scale of the observed peak without a fitted normalization. The sum for set B is closer to the data up to about \(0.24\) GeV/\(c\). At the highest observed ordinate the sums for A and B are \(3.95\times10^5\) and \(3.45\times10^5\) mb/GeV\(^2\), respectively, compared with \(2.90\times10^5\) mb/GeV\(^2\). At \(k_z\simeq0.325\) GeV/\(c\) both fall below the data (Table~\ref{tab:data}). Varying the optical parameters changes the sum much more than going from second order in \(\eta\) to the unexpanded phase.

The calculation with physical longitudinal recoil is given below for the high-momentum points. The optical profiles, projected plane wave and scalar \(S\) wave are kept fixed so that the effect of the change in kinematics can be seen directly. Appendix~\ref{app:conf} shows the separate Coulomb contributions presented at the Kyiv conference.
\section{Small longitudinal transfer and finite proton angle}\label{sec:finite}
\subsection{Longitudinal momentum transfer at small angles}
The proton angle determines \(p_\perp\), while the coherent yield is integrated over the pair momentum transfer \(\vct Q\). The internal momentum \(\vct k_\perp=\vct p_\perp-\vct Q/2\) therefore varies even at a fixed proton angle. In this section \(Q_z\) is the signed longitudinal momentum transferred to the pair in the laboratory. As in Ref.~\cite{hydrogen}, the longitudinal argument of the spectrum is \(k_z=p_z^*\) in the incident-deuteron frame, without a second boost to the recoiling pair.

For an estimate of the forward transfer, consider coherent breakup at \(Q=0\) and neglect the recoil energy of the target, which is left in its ground state. Then \(E_n=E_d-E_p^{\rm lab}\), with \(E_d=\sqrt{M_d^2+p_d^2}\), and the neutron's forward momentum is \(\sqrt{E_n^2-m_n^2-p_\perp^2}\), where \(m_n\) is the neutron mass. Writing \(\ell_z=p_d-p_z^{\rm lab}\) gives
\begin{align}
 Q_z(Q=0)&=-q_0,\qquad
 q_0=\ell_z-\sqrt{E_n^2-m_n^2-p_\perp^2}\notag\\
 &=\frac{\ell_z^2+m_n^2+p_\perp^2-E_n^2}
         {\ell_z+\sqrt{E_n^2-m_n^2-p_\perp^2}} .
 \label{eq:qmin}
\end{align}
The second form avoids the subtraction of nearly equal momenta. Approximating its denominator by \(2\ell_z\) gives the estimate used at zero angle in Ref.~\cite{Kobushkin2008}. The physical masses (\(m_p=938.272\), \(m_n=939.565\), \(M_d=1875.613\) MeV/\(c^2\)) retain the binding deficit. For \(p_\perp=0\), Eq.~\eqref{eq:qmin} gives \(q_0\simeq0.46\), \(0.56\), and \(3.0\) MeV/\(c\) at \(k_z=0\), \(0.020\), and \(0.100\) GeV/\(c\), respectively. The estimate reaches about 39 MeV/\(c\) at \(k_z=0.325\) GeV/\(c\). The fixed few-MeV transfers used below isolate their effect near the peak; the estimate~\eqref{eq:qmin} gives a second scale for comparison.
\subsection{Optical profiles and the external Coulomb pole}
To include longitudinal transfer in the strong amplitude, we use the Gaussian prescription of Ref.~\cite{hydrogen}. The impact-parameter factors of Eq.~\eqref{eq:profiles} become
\begin{align}
 \Gamma_N^{z}
 &=\overline C_N\exp[-d_N b_N^2+\ii\xi_N Q_zr_z/2],\notag\\
 \overline C_N&=C_N\ee^{-g_N Q_z^2},\qquad
 \xi_p=-1,\quad \xi_n=+1 .
 \label{eq:profilesz}
\end{align}
The prescription uses the same Gaussian width for the transverse and longitudinal momentum damping; it does not fit a separate longitudinal slope. In single scattering \(G_\pm\) is replaced by \(G_\pm^z=G((\pm\vct Q/2,\pm Q_z/2),\vct k)\). In optical double scattering the opposite longitudinal phases cancel. Coulomb terms that contain the nuclear profiles use the same factors (Appendix~\ref{app:finite}). Neutron removal still uses Eq.~\eqref{eq:loss}, since the unobserved states in this channel need not obey the recoil condition for a target in its ground state.

At finite \(p_\perp\) the pure Coulomb term is singular if one sets \(Q_z=0\). Since \(n_S(\vct q)=1+O(q^2)\), the form factor begins with \(G((-\vct Q/2,0),\vct k)=-\vct Q\cdot\nabla_\perp W(\vct k)/2+O(Q^2)\), and \(\nabla_\perp W\) need not vanish as \(\vct k_\perp\to\vct p_\perp\). The \(1/Q^2\) propagator gives a squared Born amplitude of order \(1/Q^2\); the integral with measure \(Q\dd Q\) then diverges logarithmically. A finite excitation energy supplies the physical scale at large distances where the sudden approximation fails \cite{CapelCCE}.

For a heavy target, neglect \(Q^0\) in the laboratory. In the incident-deuteron frame the exchanged energy is \(\omega=-\gamma_d v_d Q_z>0\), the longitudinal component is \(Q_z^*=\gamma_d Q_z\), and the photon denominator is \(Q^2+Q_z^2\). Expansion of the transition charge at small momentum and current conservation give the leading electric-dipole charge and current proportional to \(\ii\vct Q^*\cdot\vct D\) and \(\ii\omega\vct D\), with \(\vct D\) the transition dipole. The target moves with velocity \(-v_d\hat{\vct z}\) in that frame. Contracting the currents therefore adds \(v_d\) times the longitudinal transition current to the charge term, so the longitudinal dipole has coefficient \(Q_z^*+v_d\omega=Q_z/\gamma_d\) \cite{WintherAlder}. With the projected \(S\) wave of Eq.~\eqref{eq:ff}, the first-order pole amplitude is
\begin{align}
 \F_1^{\rm pole}
 &=\eta\,\frac{\vct Q\cdot\nabla_\perp W+
                   (Q_z/\gamma_d)\partial_{k_z}W}
                  {Q^2+Q_z^2}.
 \label{eq:e1}
\end{align}
The derivatives in the pole term are taken at \((\vct p_\perp,k_z)\), and the elastic charge form factor tends to unity there. The longitudinal suppression comes from the current term in the contraction; replacing the longitudinal momentum in the charge by \(\gamma_d Q_z\) alone does not give Eq.~\eqref{eq:e1}.

Write \(D_\perp=|\nabla_\perp W|\), \(D_z=\partial_{k_z}W\), and \(q_\parallel=|Q_z|\). Let \(\Lambda\) be the upper transverse momentum used to integrate the leading pole alone. The transverse--longitudinal cross term vanishes upon azimuthal integration. With \(y=\Lambda^2/q_\parallel^2\), the integral of the squared pole is
\begin{equation}
\begin{aligned}
 &\int_{Q<\Lambda}\dd^2Q\,|\F_1^{\rm pole}|^2 \\
 &\quad=\frac{\pi\eta^2D_\perp^2}{2}
       \left[\ln(1+y)+\frac{1}{1+y}-1\right]\\
 &\qquad+\frac{\pi\eta^2D_z^2}{\gamma_d^2}\frac{y}{1+y}.
\end{aligned}
\label{eq:softsq}
\end{equation}
The transverse term contains the radial integral of \(Q^3/(Q^2+q_\parallel^2)^2\) and an angular factor \(\pi\). For the longitudinal term they are \(Qq_\parallel^2/(Q^2+q_\parallel^2)^2\) and \(2\pi\). Since \(D_\perp=O(p_\perp)\) for an \(S\) wave, the transverse integral contains \(p_\perp^2\ln(\Lambda/q_\parallel)\); it vanishes at \(p_\perp=0\). At fixed \(k_z\ne0\) the longitudinal integral can have a finite limit as \(Q_z\to0\): the growing pole compensates for the shrinking region \(Q\sim q_\parallel\). At \(k_z=0\), \(D_z\) also vanishes. Thus taking the limit after integration need not give the spectrum defined at \(Q_z=0\). The regular transverse terms and their interference are retained in the numerical spectrum.
\subsection{Matching the long-range Coulomb amplitude}
To compare spectra at small nonzero transfer, we match the pure Coulomb term to Eq.~\eqref{eq:e1} while keeping its known transverse form away from the pole. Define $H_\eta(u)=-\ii\,\widetilde D_\eta(u)/(2\pi)$ for $u>0$, obtained from Eq.~\eqref{eq:point} and the localized finite-size transform. Put $s=\sqrt{Q^2+Q_z^2}$, $G_T=G((-\vct Q/2,0),\vct k)$, $G_L=-Q_z\partial_{k_z}W(\vct k)/(2\gamma_d)$, and $w=\exp[-(Q/Q_m)^2]$. We use the pure contribution
\begin{align}
 \F_{\eta,\rm pure}^{m}
 &=w H_\eta(s)(G_T+G_L)
     +(1-w)H_\eta(Q)G_T .
 \label{eq:joined}
\end{align}
At \(Q=0\) the second term is regular because \(1-w=O(Q^2)\). At \(Q_z=0\) the expression reduces to \(H_\eta(Q)G_T\) for every \(Q>0\). The matching imposes the one-photon dipole limit and recovers the transverse eikonal phase; higher-order longitudinal dynamics are not included. We take \(Q_m=50\) MeV/\(c\) and compare with \(Q_m\) between 25 and 100 MeV/\(c\) and with \(H_\eta(s)G_T\) in place of the interpolated transverse part. This probes the matching prescription. The expansion and the unexpanded phase use the same matching, so their difference at fixed \(Q_z\) isolates the effect of truncation.

In the spectator amplitude at nonzero \(p_\perp\), restore \(\exp(-\ii\vct p_\perp\cdot\vct r_\perp)\) in Eq.~\eqref{eq:remamp}. The inner azimuthal integral still gives a modified Bessel function, now with a complex argument (Appendix~\ref{app:finite}). For fixed nonzero \(Q_z\), azimuthal symmetry makes the spectrum even in \(p_\perp\); with a smooth kernel the first change from the on-axis value is quadratic. Equation~\eqref{eq:softsq} explains why that coefficient need not remain bounded as \(Q_z\to0\).

\subsection{Variation of the spectrum near zero angle}
\begin{figure*}[t]
 \centering\includegraphics[width=\textwidth]{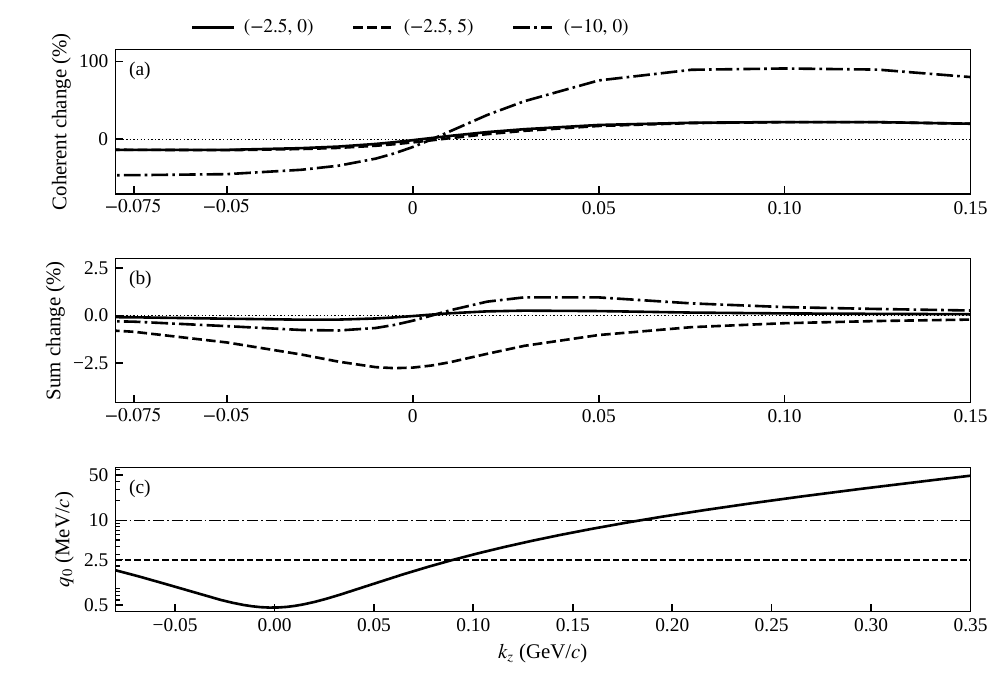}
 \caption{Sensitivity to small momentum transfers. (a,b) Changes relative to $Q_z=p_\perp=0$ for the coherent yield and the sum, for optical set A. Momentum pairs in the legend give $(Q_z,p_\perp)$ in MeV/$c$. (c) Laboratory deficit $q_0$ from Eq.~\eqref{eq:qmin}; the horizontal lines mark $2.5$ and $10$ MeV/$c$. The fixed-transfer curves do not use the rising deficit in panel (c).}
 \label{fig:finite}
\end{figure*}

Table~\ref{tab:finite} lists changes from the result at \(Q_z=p_\perp=0\). At \(k_z=20\) MeV/\(c\), \(Q_z=-2.5\) MeV/\(c\), and \(p_\perp=0\), the coherent yield rises by about \(9\%\) for A and \(8\%\) for B, against \(0.2\%\) and \(0.6\%\) in the sum. In set A the strong amplitude alone gives a rise of about \(9\%\) here. The squared longitudinal dipole contributes about 30 mb/GeV\(^2\), already included with the interference in the table. At the peak the coherent change has the opposite sign. Because the coherent yield is invariant under simultaneous reversal of \(k_z\) and \(Q_z\), a fixed negative \(Q_z\) can make the shape asymmetric (Fig.~\ref{fig:finite}). At \(k_z=0\) and \(Q_z=-2.5\) MeV/\(c\), changing \(p_\perp\) from zero to \(0.032\) MeV/\(c\) changes the sum by less than \(0.001\%\). This is the geometric scale for a proton angle of 7 microradians at about \(p_d/2\), used here as an example rather than an acceptance for the carbon measurements. At \(p_\perp=5\) MeV/\(c\) the sum near the peak falls by about \(3\%\), mainly because the spectator spectrum changes.

For fixed \(|Q_z|=2.5\) and 10 MeV/\(c\) and \(p_\perp\le5\) MeV/\(c\), the second-order result differs from the unexpanded coherent yield by at most \(0.03\%\) at the calculated points between \(k_z=-0.10\) and 0.325 GeV/\(c\). At four momenta between 0 and 0.325 GeV/\(c\), changing the matching of the pure Coulomb term changes the sum by less than \(0.04\%\) in the cases tested, while the coherent yield can change by up to \(3\%\). Refining the quadratures and extending \(Q_{\max}\) to 1 GeV/\(c\) changes each tested yield by less than \(10^{-6}\) in relative terms. Setting \(Q_z=p_\perp=0\) recovers the transverse result, and a common shift of the transverse phase leaves the squared amplitudes invariant within the quadrature accuracy.

Figure~\ref{fig:finite} shows the changes around the peak and the increase of \(q_0\) with \(k_z\). Holding \(Q_z=-q_0(k_z)\) fixed in each integral gives the estimates at \(k_z=0\), \(0.020\), and \(0.100\) GeV/\(c\). They indicate the scale set by opening the breakup channel. Farther from the peak the transfer must instead be calculated for the measured proton momentum and the recoil.

To compare with bins in the measured spectrum, energy conservation fixes \(Q_z\) at each \(\vct Q\) in the coherent integral. The neutron momentum components are \(\vct Q-\vct p_\perp\) and \(p_d+Q_z-p_z^{\rm lab}\). For a target of mass \(M_A\) the recoil energy is \(T_A=\sqrt{M_A^2+Q^2+Q_z^2}-M_A\), and \(E_d=E_p^{\rm lab}+E_n+T_A\) selects \(Q_z\) on the physical branch. The predicted bin count is obtained by folding the spectrum with the angle and momentum response. If \(\mathcal R_i(\vct p^{\,\rm lab})\) is the probability of assigning the proton to bin \(i\) and \(\mathcal L\) the integrated luminosity, then
\begin{equation}
 {\cal N}_i=\mathcal L\int
  \frac{\dd^3p^{\,\rm lab}}{E_p^{\rm lab}}\,
   \mathcal R_i(\vct p^{\,\rm lab})\,\I(k_z,p_\perp).
 \label{eq:acceptance}
\end{equation}
The published graph does not specify \(\mathcal R_i\), so Table~\ref{tab:finite} compares differential yields without averaging over the bins.
\begin{table*}[t]
 \setlength{\belowcaptionskip}{1.5pt}
 \caption{Relative changes (\%) in the coherent yield and the two-channel sum, with respect to $Q_z=p_\perp=0$. The Coulomb phase is retained in the continuation~\eqref{eq:joined}. Momenta are given in MeV/$c$. The last row uses the forward estimate \eqref{eq:qmin} at each indicated momentum.}
 \centering
 \setlength{\tabcolsep}{5pt}
\begin{tabular}{@{}rr*{8}{r}@{}}\toprule
\multicolumn{2}{c}{} & \multicolumn{4}{c}{A} & \multicolumn{4}{c}{B}\\
\cmidrule(lr){3-6}\cmidrule(lr){7-10}
& & \multicolumn{2}{c}{$k_z=0$} & \multicolumn{2}{c}{$k_z=20$} & \multicolumn{2}{c}{$k_z=0$} & \multicolumn{2}{c}{$k_z=20$}\\
\cmidrule(lr){3-4}\cmidrule(lr){5-6}\cmidrule(lr){7-8}\cmidrule(lr){9-10}
$Q_z$ & $p_\perp$ & $\I_{\coh}$ & $\I_{\rm sum}$ & $\I_{\coh}$ & $\I_{\rm sum}$ & $\I_{\coh}$ & $\I_{\rm sum}$ & $\I_{\coh}$ & $\I_{\rm sum}$\\\midrule
$-2.5$ & $0$ & $-1.1$ & $-0.03$ & $+9.2$ & $+0.22$ & $-0.58$ & $-0.06$ & $+7.6$ & $+0.64$\\
$-2.5$ & $1$ & $-1.3$ & $-0.14$ & $+9.1$ & $+0.12$ & $-0.73$ & $-0.17$ & $+7.4$ & $+0.55$\\
$-2.5$ & $5$ & $-4.0$ & $-2.8$ & $+6.6$ & $-2.0$ & $-4.0$ & $-2.9$ & $+4.4$ & $-1.7$\\
$-10$ & $0$ & $-9.9$ & $-0.28$ & $+31.2$ & $+0.73$ & $-6.3$ & $-0.59$ & $+28.2$ & $+2.4$\\
$-q_0(k_z)$ & $0$ & $-0.07$ & $-0.002$ & $+2.8$ & $+0.06$ & $-0.03$ & $-0.003$ & $+1.9$ & $+0.16$\\
\bottomrule
\end{tabular}
 \label{tab:finite}
\end{table*}

\subsection{Forward recoil on the high-momentum tail}
To test the spectrum in the region of the excess, we also calculate the yield at five measured momenta between \(0.305\) and \(0.345\) GeV/\(c\). In the coherent integral \(Q_z\) is determined at each \(\vct Q\) by energy conservation with the target and neutron on shell, as in Eq.~\eqref{eq:sit-recoil}. We retain the longitudinal phase-space factor of Eq.~\eqref{eq:hjac}, with the carbon mass in place of the hydrogen target mass. In the optical and dipole amplitudes we use the prescription of this section, and for neutron removal we retain the spectator spectrum. At \(Q=0\) the magnitude \(|Q_z|\) grows from about 33 to 47 MeV/\(c\) over the five points. These transfers are much larger than the few-MeV values used to examine the peak.

At \(k_z\simeq0.325\) GeV/\(c\), recoil raises the two-channel sum by about \(2\%\) in set A and \(10\%\) in B relative to the transverse result. The resulting yields are \(3.6\) and \(2.8\) mb/GeV\(^2\), well below the published value of \(18.5\) mb/GeV\(^2\) [Fig.~\ref{fig:data}, lower panel]. At all five points the Coulomb change of the sum at \(p_t=0\) is below \(0.08\%\). Giving the proton \(p_t=5\) MeV/\(c\) changes the on-shell sum by less than \(0.3\%\) there. The excess refers to the difference between the measured and calculated yields; it need not be a local maximum of the measured spectrum. The result limits the role of the mechanisms included here. A quantitative account of the remaining yield would require the spin and \(D\)-wave terms, the \(pn\) continuum, and the measured acceptance in the high-momentum region.

\section{Carbon and hydrogen spectra at 3.5 GeV/\(c\)}\label{sec:sitnik}
\subsection{Kinematics of the Strela measurement}
Sitnik measured inclusive proton spectra on carbon and polyethylene with the Strela spectrometer at \(p_d=3.5\) GeV/\(c\) \cite{Sitnik2019}. The longitudinal variable \(q_L\) is the proton momentum in the rest frame of the incident deuteron, \(p_z^*\) in Eq.~\eqref{eq:boost}. For the first comparison we interpret the published cuts as limits on the detected proton's transverse momentum \(p_t=|\vct p_\perp|\), which is unchanged by the boost along the beam. Since the final \(pn\) pair was not reconstructed, a cut on \(p_t\) still admits different values of the internal momentum \(\vct k_\perp=\vct p_\perp-\vct Q/2\) in the coherent integral. Basilev \textit{et al.} \cite{Basilev2015} reported the same data and extracted the hydrogen yield by subtracting the carbon contribution from polyethylene. The older forward data shown for comparison do not cover the five transverse selections of Ref.~\cite{Sitnik2019}.
At this beam momentum \(\eta\simeq0.050\) for carbon and \(\gamma_d\simeq2.12\). The longitudinal transfer \(Q_z\) varies with the recoil in the coherent channel. For a target of mass \(M_A\) left in its ground state, the neutron carries transverse momentum \(\vct Q-\vct p_\perp\) and longitudinal momentum \(\ell_z+Q_z\), where \(\ell_z=p_d-p_z^{\rm lab}\). Energy conservation in the laboratory then gives
\begin{equation}
 \begin{aligned}
 E_d-E_p^{\rm lab}
 &=E_n+T_A,\\
 E_n&=\sqrt{m_n^2+|\vct Q-\vct p_\perp|^2+(\ell_z+Q_z)^2},\\
 T_A&=\sqrt{M_A^2+Q^2+Q_z^2}-M_A.
 \end{aligned}
 \label{eq:sit-recoil}
\end{equation}
We take the solution that joins the forward-neutron branch of Eq.~\eqref{eq:qmin}. Since the energy transfer to the pair is \(Q^0=-T_A\), the argument \(s\) in Eq.~\eqref{eq:joined} becomes \(\sqrt{Q^2+Q_z^2-T_A^2}\). The boost in Eq.~\eqref{eq:boost} gives \(Q_z^*+v_d Q^{*0}=Q_z/\gamma_d\) also with recoil, so the dipole numerator is unchanged. At \(q_L=Q=0\), the magnitude \(|Q_z|\) is \(1.2\) MeV/\(c\) on the axis and rises to \(2.1\), \(4.9\), and \(16\) MeV/\(c\) at \(p_t=40\), \(80\), and \(160\) MeV/\(c\), respectively. Thus a transverse cut affects both the transition density and the distance from the Coulomb pole. We use this on-shell kinematics in the optical amplitudes of Sec.~\ref{sec:finite}, with the leading eikonal measure \(\dd^2Q\) for the carbon recoil. The treatment of the electromagnetic dynamics is still the one given in that section. Neutron removal is calculated inclusively in the spectator approximation.
\subsection{Spectra integrated over the transverse cuts}
For a cut with upper limit \(p_c\) we first take the selections to be nested disks \(p_t<p_c\), with a common yield scale within each of the figures in Ref.~\cite{Sitnik2019}. With uniform transverse acceptance the calculated distribution in \(q_L\) is
\begin{equation}
 \begin{aligned}
 Y_C(q_L;p_c)&=2\pi\int_0^{p_c}p_t\,\dd p_t\,
 \frac{\mathcal I_C(q_L,p_t)}{E_p^*},\\[-1pt]
 E_p^*&=\sqrt{m_p^2+q_L^2+p_t^2}.
 \end{aligned}
 \label{eq:sit-yield}
\end{equation}
The invariant spectrum \(\mathcal I_C\) includes coherent breakup and neutron removal. The factor \(1/E_p^*\) converts it to a momentum density in the incident-deuteron frame. This integrated yield is distinct from the invariant spectrum averaged over an aperture. We integrate over the successive annuli bounded by 10, 20, 40, 80, and 160 MeV/\(c\) and add the parts inside \(p_c\). Reference~\cite{Basilev2015} gives the geometric acceptance as a function of laboratory momentum, without the joint response in \(q_L\) and \(p_t\). The ratios below therefore assume uniform transverse acceptance. The carbon spectra given in arbitrary units are used only to compare relative yields.
\subsection{Elementary scattering inputs at $p_d=3.5$ GeV/$c$}
Near the spectator peak a nucleon of the incident deuteron has momentum close to \(p_d/2\), corresponding to a kinetic energy of about 1.05 GeV. The proton profile for carbon is fixed by \(p+{}^{12}\)C elastic scattering at 1.04 GeV \cite{Bertini1973}; the neutron profile uses total and nonelastic \(n+\)C cross sections on either side of that energy \cite{Schimmerling1973}. No breakup spectra enter these estimates. Write \(x=-t\geq0\) for the squared invariant momentum transfer and \(B_N\) for the logarithmic slope of the elementary strong differential cross section. With \(\beta_N\) in fm we have \(B_N=\beta_N^2/(\hbar c)^2\). For cross sections in mb and \(x\) in GeV\(^2\), the optical theorem and Eq.~\eqref{eq:profiles} give
\begin{equation}
 \begin{aligned}
 \frac{\dd\sigma_N^{\rm el}}{\dd x}&=A_N\ee^{-B_N x},\\
 A_N&=\frac{\sigma_N^2(1+\rho_N^2)}{16\pi C_0},\qquad
 C_0=0.3894\ {\rm mb\,GeV}^2.
 \end{aligned}
 \label{eq:elemstrong}
\end{equation}
For the proton we fit the interval \(0.005<x<0.025\) GeV\(^2\) of EXFOR O0125002, using \(\dd\sigma/\dd x=(\pi/k_{\rm cm}^2)\dd\sigma/\dd\Omega_{\rm cm}\) with centre-of-mass momentum \(k_{\rm cm}\). The fit includes the interference of the finite-charge Rutherford and strong amplitudes. For \(\rho_p=0\) it gives \(\sigma_p\simeq348\) mb and \(B_p\simeq63\) GeV\(^{-2}\) (\(\beta_p\simeq1.57\) fm). The eight data points give \(\chi^2=2.7\) for six degrees of freedom with the reported point errors, excluding the \(7\%\) common normalization uncertainty. Wider fitting intervals and variations of the real part test the sensitivity to these choices. For the neutron, linear interpolation of the two EXFOR 10082 data series gives \(\sigma_n\simeq353\) mb and \(\sigma_n^{\rm nonel}\simeq267\) mb. For \(\rho_n=0\), the Gaussian elastic integral \(A_n/B_n=\sigma_n-\sigma_n^{\rm nonel}\) then gives \(\beta_n\simeq1.69\) fm. The natural-carbon neutron data serve as a proxy for \(^{12}\)C. These profiles are used at 3.5 GeV/\(c\); the zero-angle calculation at 9.1 GeV/\(c\) uses sets A and B.

\begin{table*}[t]
 \setlength{\belowcaptionskip}{1.5pt}
 \centering
 \caption{Carbon peak heights in percent of the peak at $p_c=80$ MeV/$c$. The two estimates from Ref.~\cite{Sitnik2019} are normalized separately. A and B are the reference optical sets; the constrained profiles use elementary scattering at the lower beam energy. All calculated columns use the unexpanded phase.}
 \label{tab:sitnik}
 \begin{tabular}{@{}cccccc@{}}
 \hline\hline
 $p_c$ (MeV/$c$)& Fig.~3 (largest bin) & Fig.~4 (maximum) & A & B & Constrained\\
 \hline
 10 & 10 & 15 & 7 & 7 & 7 \\
 20 & 33 & 33 & 24 & 25 & 25 \\
 40 & 74 & 76 & 62 & 63 & 64 \\
 80 & 100 & 100 & 100 & 100 & 100 \\
 160& 101 & 100 & 115 & 116 & 117 \\
 \hline\hline
 \end{tabular}
\end{table*}
\begin{figure*}[t]
 \centering\includegraphics[width=\textwidth]{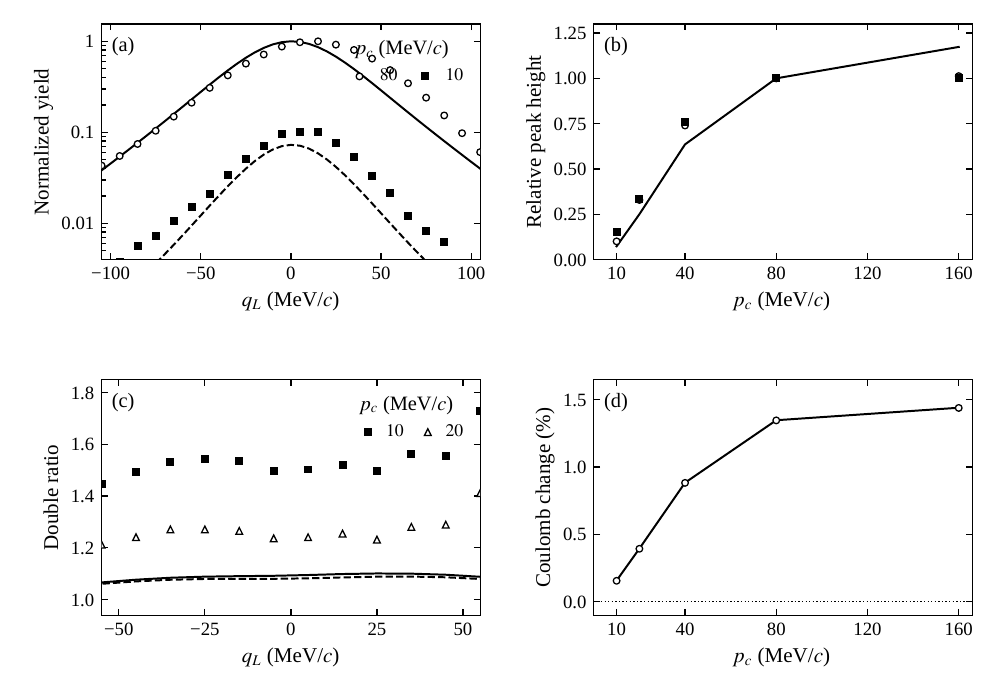}
 \caption{Transverse cuts at $p_d=3.5$ GeV/$c$. (a) Spectra for $p_c=80$ and 10 MeV/$c$ (points: Fig.~3 of Ref.~\cite{Sitnik2019}; solid and dashed curves: calculation). Each pair uses its 80 MeV/$c$ peak as the scale. (b) Relative carbon peak heights: curve, calculated; circles and squares, respectively, from Figs.~3 and 4 of the same source. (c) Double ratio $D$ from Fig.~2 of that source (squares: 10; triangles: 20 MeV/$c$) and the calculation (solid and dashed lines, respectively). (d) Coulomb change of the inclusive carbon yield at $q_L=0$: solid curve, unexpanded phase; open circles, complete quadratic order; dotted line, linear result.}
 \label{fig:sitnik}
\end{figure*}
\subsection{Proton recoil and elastic \(np\) strength on hydrogen}\label{sec:H}
The hydrogen yield is extracted from carbon and polyethylene as \(Y_H=(Y_{\mathrm{CH}_2}-Y_C)/2\), with \(Y_{\mathrm{CH}_2}\) per molecule \cite{Basilev2015}. For a proton target the same deuteron wave function and operator \(1-S_pS_n\) apply, now with separate \(pp\) and \(np\) profiles. The charge is \(Z=1\) and \(\eta_H\simeq0.0083\). We use a Gaussian charge distribution of width \(a_H=\sqrt{2/3}\,r_p\simeq0.69\) fm, set by the proton rms charge radius \(r_p=0.84\) fm \cite{Mohr2025}; the strong widths are determined separately.
In Eq.~\eqref{eq:sit-recoil} we set \(M_A=m_p\) and choose the branch with a forward neutron at small transfer. The recoil matters for a light target. At \(q_L=p_t=0\), \(|Q_z|\) is about \(1.2\), \(3.4\), and 22 MeV/\(c\) for \(Q=0\), 50, and 150 MeV/\(c\), respectively. At \(Q=0\) it increases to about \(4.9\) MeV/\(c\) when \(p_t=80\) MeV/\(c\). Let \(E_n\) and \(E_R\) be the neutron and recoil-proton energies and \(p_{nz}=\ell_z+Q_z\). The longitudinal phase-space Jacobian, normalized to the leading eikonal measure, is
\begin{equation}
 J_H=\frac{m_p\ell_z}{|E_R p_{nz}+E_n Q_z|}.
 \label{eq:hjac}
\end{equation}
Eliminating \(Q_z\) with the energy delta function brings in the derivative \(p_{nz}/E_n+Q_z/E_R\). Together with \(1/(E_n E_R)\) this gives the denominator of Eq.~\eqref{eq:hjac}; the numerator is the small-recoil reference. At \(Q_{\max}\) the neutron and recoil proton have the same longitudinal velocity. The substitution \(Q=Q_{\max}\sin u\) for each azimuth removes the integrable endpoint singularity of \(J_H\). The photon propagator has the same invariant virtuality as on carbon, the Gaussian convolutions retain their transverse arguments, and the boost of \(Q_z\) is included in the internal kinematics.
For comparison, the K2H parameters of Ref.~\cite{hydrogen} are \(\sigma_p=\sigma_n=65\) mb, \(\beta_p=\beta_n=0.75\) fm and \(\rho_N=10^{-3}\). The corresponding nonelastic Gaussian integral, about 50 mb, exceeds the free \(np\) nonelastic cross section. We use elementary \(pp\) and \(np\) total cross sections of about 47 and 39 mb at \(p_d/2\) \cite{PDG2024}. We take \(B_p=5.6\) GeV\(^{-2}\) and obtain \(B_n=5.7\) GeV\(^{-2}\) from the four forward bins of Ref.~\cite{Perl1970}, keeping their forward normalization and averaging the calculation over each bin. The values \(-0.18\) and \(-0.50\) for the ratios of real to imaginary parts are based on neighbouring measurements from the same database. They are varied together with the slopes. In the strong Gaussians \(x\) replaces \(Q^2\). The exact boost gives \(Q_z^*=\gamma_d(Q_z+v_dT_A)\); sharing the recoil between the two nucleons to leading order gives \(k_z=q_L-Q_z^*/2\). The direct neutron term is then evaluated with the observed proton momentum before the collision. The scalar model does not fully antisymmetrize the two protons in \(dp\to ppn\).

The Gaussian \(np\) amplitude describes the forward cone. To account for the measured elastic strength outside it, we use coherent recoil only for \(x<x_m\) and divide the elementary cross section into
\begin{equation}
 \begin{aligned}
 \sigma_{np}^{\rm fwd}(x_m)&=\frac{A_n}{B_n}
                          \bigl(1-\ee^{-B_n x_m}\bigr),\\
 \sigma_{np}^{\rm out}&=\sigma_{np}^{\rm el}-\sigma_{np}^{\rm fwd},\\
 \sigma_{np}^{\rm nonel}&=\sigma_{np}^{\rm tot}-\sigma_{np}^{\rm el}.
 \end{aligned}
 \label{eq:matchH}
\end{equation}
At \(x_m=0.30\) GeV\(^2\) the three terms are about 14, 11, and 14 mb, respectively. Inside the cone the coherent cross section is integrated with \(J_H\). Outside it elastic neutron scattering is counted with the spectator contribution. The last two terms use the shape of Eq.~\eqref{eq:loss} scaled to their free cross sections; the forward cone is counted only in the coherent part. This partition preserves the sum of the elementary strengths and treats the observed fast proton as the projectile spectator also for large-transfer \(np\) events. Target-proton feedthrough, spin, and three-body final-state interactions are not included. We vary \(x_m\) between 0.20 and 0.45 GeV\(^2\) and keep the NN reference, which uses free total cross sections with forward diffraction and nonelastic neutron removal only. None of these parameters is fitted to a breakup peak.

\begin{figure*}[t]
 \centering\includegraphics[width=\textwidth]{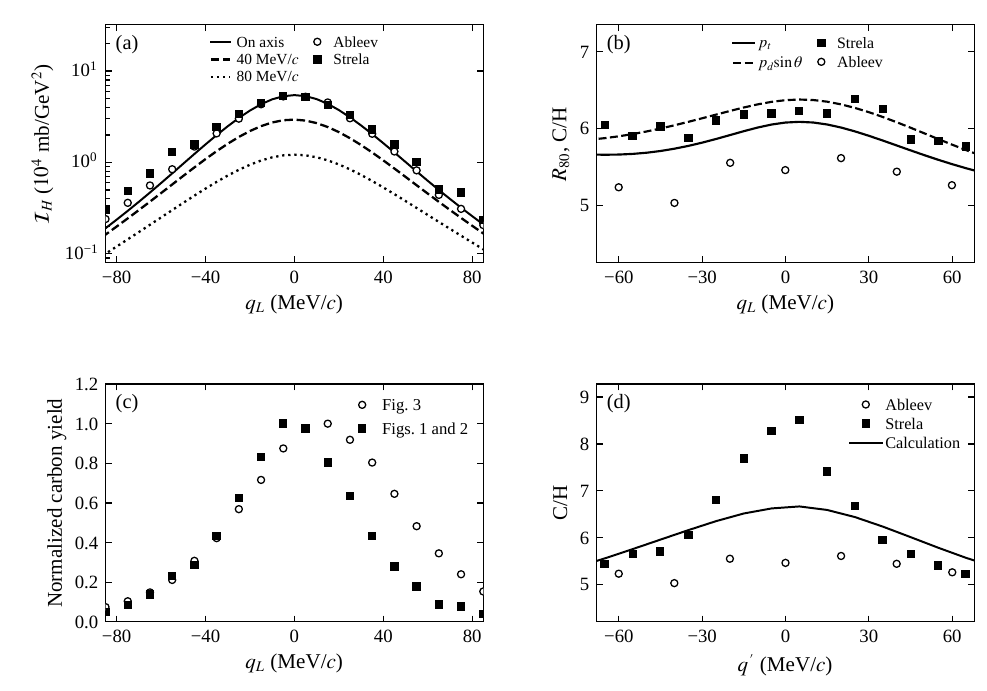}
 \caption{Comparison with the published spectra. (a) The hydrogen data at 3.5 and 9 GeV/\(c\) from Ref.~\cite{Sitnik2019}. The calculated spectrum at 3.5 GeV/\(c\) is shown on the axis and averaged over transverse disks of approximately 40 and 80 MeV/\(c\) at the peak. (b) The measured \(R_{80}\), the older forward ratio reproduced in Ref.~\cite{Sitnik2018}, and the calculations for the two cut conventions. (c) Carbon shapes: circles, Fig.~3 of Ref.~\cite{Sitnik2019}; squares, the product of the ordinates in Figs.~1 and 2 divided by \(E_p^*\) at \(p_t=0\). Each is scaled to its own maximum. (d) The two data sets in Fig.~6 of Ref.~\cite{Basilev2015}, compared with the calculation using the same variable \(q'\). Open symbols in (a), (b), and (d) identify the older measurements.}
 \label{fig:proton}
\end{figure*}
\subsection{Comparison with the measured spectra}
The hydrogen spectra of Strela and Ableev \textit{et al.} \cite{Sitnik2019,Ableev1992}, at 3.5 and 9 GeV/\(c\), both peak near \(5.3\times10^4\) mb/GeV\(^2\) [Fig.~\ref{fig:proton}(a)]. Their full widths at half maximum are about 63 and 57 MeV/\(c\), respectively. In the six common bins with \(|q_L|\le25\) MeV/\(c\), the ratio of the newer ordinates to the 1992 values has a median of 1.02 and a range of 0.94 to 1.12. Normalizing by that median still leaves a ratio of about 1.18 at \(q_L=+45\) MeV/\(c\). Thus the measurements support the same peak scale more closely than they support a common shape. The calculated widths for the 80 MeV/\(c\) disk are about 72 and 73 MeV/\(c\).

The six central ordinates in Fig.~5 of Ref.~\cite{Basilev2015} agree with those in Fig.~1 of Ref.~\cite{Sitnik2019} to the accuracy of digitization, although the descriptions give different upper transverse limits. They are presentations of the same data. The independent zero-angle measurement of Ref.~\cite{Perdrisat1987} covers higher internal momenta and does not determine the scale at the peak.

At fixed \(q_L\), write \(R(p_c)=Y_C/Y_H\) and \(D(p_c)=R(p_c)/R(80)\), with \(p_c\) in MeV/\(c\).

The conference abstract~\cite{Sitnik2018} gives \(p_d\sin\theta\), while Ref.~\cite{Sitnik2019} does not repeat the definition. The six central ratios agree between the publications for each cut to the accuracy of digitization. If \(\theta\) is the laboratory angle of the outgoing proton, then \(p_t=p_p^{\rm lab}\sin\theta\); near the peak \(p_p^{\rm lab}\simeq p_d/2\). A cut labelled 80 MeV/\(c\) thus corresponds to roughly 40 MeV/\(c\) in \(p_t\) under the conference convention. In the calculation we use \(p_p^{\rm lab}\) at each node. At \(q_L=0\) the hydrogen calculation gives \(5.5\times10^4\) mb/GeV\(^2\) on the axis, or about \(2.9\times10^4\) and \(1.2\times10^4\) when averaged over disks of 40 and 80 MeV/\(c\). Which quantity should be compared with Fig.~\ref{fig:proton}(a) depends on how the published cross section was corrected for acceptance.

With cuts on \(p_t\), the central six-bin medians are \(R_{80}\simeq6.0\), \(D_{10}\simeq1.09\), and \(D_{20}\simeq1.08\); with cuts on \(p_d\sin\theta\) they are \(6.3\), \(1.05\), and \(1.04\). The published ordinates give \(6.2\), \(1.51\), and \(1.25\). The six values of \(D_{10}\) lie between about 1.50 and 1.54, so the stronger rise appears across the central interval. The nested selections share events, and this range is no measure of statistical uncertainty. The three central 1992 points reproduced in Ref.~\cite{Sitnik2018} give about 5.6 for the forward \(C/H\) ratio; they provide a scale comparison without repeating the narrow cuts. Varying the elementary profiles and the \(np\) matching boundary gives \(D_{10}\simeq1.07\)--\(1.12\) for cuts on \(p_t\). A Gaussian representation of the tabulated Nijmegen-I \(S\) wave~\cite{NNOnline} gives \(D_{10}\simeq1.09\) with the same profiles. A Gaussian smearing with longitudinal width of 7.5 MeV/\(c\) leaves the central double ratios nearly unchanged.
Basilev \textit{et al.}~\cite{Basilev2015} plotted the ratio for the same new run against the signed magnitude \(q'=\operatorname{sgn}(q_L)\sqrt{q_L^2+p_t^2}\). We integrate the calculation over the same 10 MeV/\(c\) bins in \(q'\). At fixed \(p_t\), the Jacobian is \(|q'|/\sqrt{q'^2-p_t^2}\) for \(p_t<|q'|\). With uniform acceptance the central median is 6.6, against 7.5 for the new measurements [Fig.~\ref{fig:proton}(d)]. This projection of the same events gives a useful comparison in another variable.

As \(p_c\) increases from 10 to 80 MeV/\(c\), the calculated carbon peak widens from about 55 to 68 MeV/\(c\); the widths read from Fig.~3 of Ref.~\cite{Sitnik2019} increase from about 62 to 83 MeV/\(c\). The calculated height for the narrow cut is about \(7\%\) of the height for the wide cut, compared with \(10\)--\(15\%\) from the two published estimates (Table~\ref{tab:sitnik}). At \(q_L=+45\) MeV/\(c\) the direct 80 MeV/\(c\) carbon plot gives 0.65 of its peak. Multiplying the hydrogen ordinate by \(R_{80}\), converting the invariant measure, and scaling to the peak gives 0.28 [Fig.~\ref{fig:proton}(c)]. A scale factor cannot change these relative heights. To interpret both as estimates of the same binned yield would require the corrections used to obtain them.

At \(q_L=0\) and \(p_c=80\) MeV/\(c\), holding \(Q_z\) at its \(Q=0\) value in the coherent integral changes the sum by about \(-1.4\%\) on hydrogen and \(+0.3\%\) on carbon. Keeping recoil but setting \(J_H=1\) reduces the hydrogen sum by about \(4\%\). Depending on the cut, the full phase raises the coherent carbon peak by \(6\)--\(16\%\). The inclusive change reaches about \(1.4\%\) on carbon and \(0.3\%\) on hydrogen. Over the central \(q_L\) interval the complete second-order calculation is within \(0.01\%\) of the unexpanded inclusive result. The effects of recoil and acceptance are therefore more relevant to this comparison than higher orders of the Coulomb phase.
\section{Conclusions}
The high-momentum excess on carbon persists when recoil is included at \(p_d=9.1\) GeV/\(c\). At \(k_z\simeq0.325\) GeV/\(c\), the kinematically allowed longitudinal transfer raises the calculated cross section to \(2.8\)--\(3.6\) mb/GeV\(^2\), depending on the optical parameters. The published spectrum gives \(18.5\) mb/GeV\(^2\) at that momentum. At the five points between \(0.305\) and \(0.345\) GeV/\(c\), the Coulomb correction to the sum with recoil is below \(0.08\%\); changing \(p_t\) from zero to \(5\) MeV/\(c\) changes the sum by less than \(0.3\%\). These corrections refine the calculation without accounting for the excess.

Near the main peak the Coulomb effect is most apparent in coherent breakup. At zero angle the peak of the coherent cross section increases by \(12\%\) in set A and \(3.6\%\) in set B. The larger spectator yield decreases slightly and partly offsets this increase. In the transverse approximation the total change is therefore less than \(0.2\%\) at the 44 comparison points. Through second order in the Sommerfeld parameter, the squared linear amplitude and the interference of the quadratic amplitude with the strong amplitude together reproduce the coherent cross section from the unexpanded phase within \(0.1\%\) at those points.

At 3.5 GeV/\(c\) the calculated carbon-to-hydrogen ratio is on the scale of the data for the transverse cut at 80 MeV/\(c\). Reducing the cut to 10 MeV/\(c\) increases the ratio by a factor of \(1.05\)--\(1.09\) in the calculation and \(1.51\) in the published data. This difference near the main peak makes the definition of the cut and the spectrometer response important to the comparison.

A quark contribution to the high-momentum excess remains possible. To distinguish it from mechanisms omitted here, the effects of the spin-dependent amplitudes and the \(pn\) continuum should be examined with the same acceptance. Polarization measurements or the detection of both outgoing nucleons would provide a test beyond the inclusive proton spectrum.

\textit{Use of generative AI.} \mbox{ChatGPT} (OpenAI) assisted with language editing, translation, checking the algebra, and developing the scripts for quadrature, parameter fitting, digitization of published graphs, and numerical comparisons. The equations, inputs, and numerical checks are documented in the manuscript and its accompanying files. Responsibility for the scientific content rests with the authors.
\appendix
\section{Gaussian coefficients and convolutions}\label{app:gauss}
For the Fourier convention of Eq.~\eqref{eq:ff}, let \(\nu_j=1/(4\lambda_j)\), \(\Lambda_{ij}=\lambda_i+\lambda_j\), and \(\xi_{ij}=1/(4\Lambda_{ij})\). The Gaussian transforms give
\begin{align}
  W(\vct k)&=\sum_j w_j\ee^{-\nu_j k^2},&
 w_j&=\frac{A_j}{2\sqrt{\pi}N_r}
         \left(\frac{\pi}{\lambda_j}\right)^{3/2},\notag\\
 n_S(\vct q)&=\sum_{i,j}n_{ij}\ee^{-\xi_{ij}q^2},&
 n_{ij}&=\frac{A_iA_j}{4\pi N_r^2}
        \left(\frac{\pi}{\Lambda_{ij}}\right)^{3/2}.
 \label{eq:wn}
\end{align}
The normalization \(N_r^2=(4\pi)^{-1}\sum_{ij}A_iA_j(\pi/\Lambda_{ij})^{3/2}\) gives \(\sum_{ij}n_{ij}=1\). Table~\ref{tab:k2} lists the coefficients for the K2 interaction \cite{K2}. The coefficient file gives more digits because some of the Gaussian terms partially cancel.
\begin{table}[t]
 \setlength{\belowcaptionskip}{1.5pt}
 \caption{Gaussian coefficients of the K2 \(S\) wave [Eq.~\eqref{eq:wave}]. The calculation uses the unrounded coefficients from the supplementary file.}
 \label{tab:k2}
 \centering
 \begin{tabular}{@{}rrr@{}}
 \toprule
 $j$ & $A_j$ (fm$^{-3/2}$) & $\lambda_j$ (fm$^{-2}$)\\\midrule
 1 & 0.40344 & 0.32020\\
 2 & 0.046775 & 0.039446\\
 3 & $-0.72389$ & 1.7986\\
 4 & $-0.075623$ & 10.145\\
 5 & 0.33837 & 1.2472\\
 6 & 0.15249 & 0.14377\\
 7 & 0.0046576 & 0.0089672\\
 8 & 0.27054 & 6.1863\\
 9 & $-0.27708$ & 0.25781\\
 10 & 0.0078405 & 0.014815\\
 11 & $-0.13233$ & 2.7861\\
 12 & 0.00025405 & 0.0029505\\
 \bottomrule
 \end{tabular}
\end{table}
At \(\vct p_\perp=0\), the first term of \(G\) in \(\Phi\) has \(\vct k+\vct q=(-\vct\ell,k_z)\). Its \(j\)th Gaussian has the prefactor \(w_j\ee^{-\nu_j k_z^2-g_n Q^2}\), width \(a=g_n+\nu_j\), and vector \(\vct L=2g_n\vct Q\). The bound-state projection gives prefactors \(-W(\vct k)n_{ij}\ee^{-(g_n+\xi_{ij}/4)Q^2}\), widths \(a=g_n+\xi_{ij}\), and vectors \(\vct L=(2g_n+\xi_{ij})\vct Q\). In the strong double-scattering term, multiplication by \(\ee^{-g_p\ell^2}\) gives the integral \(\pi\ee^{L^2/[4(a+g_p)]}/(a+g_p)\). In Eq.~\eqref{eq:moment} each term keeps its prefactor and sign. The terms are combined at the amplitude level before squaring.
\section{Analytic first-order moments}\label{app:moments}
The first-order point-charge transform is a distribution. In the phase convention of Eq.~\eqref{eq:phase}, it is the limit of \(-4\pi/(\ell^2+\mu^2)+4\pi^2 d_\mu\delta^{(2)}(\vct\ell)\) as \(\mu\to0^+\), where \(\mu\) is the screening wavenumber and
\begin{align}
 d_\mu&=c_C-2\ln(\mu b_0/2)-2\gamma_E ,\notag\\
 \int\frac{\ee^{-a\ell^2}\dd^2\ell}{\ell^2+\mu^2}
 &=\pi\ee^{a\mu^2} E_1(a\mu^2).
 \label{eq:screen}
\end{align}
The substitution \(\ell^2\) for the radial variable gives the second line. On the Gaussian \(\ee^{-a\ell^2}\) the singular and zero-mode terms combine to give \(4\pi^2[c_C+\ln(4a/b_0^2)-\gamma_E]\). With a nonzero linear vector \(\vct L\) the angular average gives \(I_0(L\ell)\); \(I_0(L\ell)-1=O(\ell^2)\) makes the extra integral convergent with \(\mu=0\). The form factor of Eq.~\eqref{eq:charge} leaves the zero-mode coefficient unchanged because \(F_{\rm ch}(0)=1\). At \(Q>0\) the transform is \(\widetilde t(Q)=-4\pi F_{\rm ch}(Q)/Q^2\).

For real $z\ne0$ write $\E(z)=\Ei(z)-\gamma_E-\ln|z|$, and set $\E(0)=0$. Equivalently $\E(z)=\int_0^1(\ee^{uz}-1)\dd u/u$. Let $D=L^2/4$, $h=a_0^2/4$, $H=a+h$. After the angular average, the subtracted moment is
\begin{align}
 \mathcal M_t(a,L)=4\pi^2\Big[&
   c_C+\ln(4H/b_0^2)-\gamma_E-\E(D/H)\notag\\
  &+\frac{\kappa a_0^2}{2H}\ee^{D/H}\Big].
 \label{eq:mt}
\end{align}
For instance, termwise integration of \(I_0(L\ell)-1\) with weight \(\ee^{-H\ell^2}\dd\ell/\ell\) gives \(\tfrac12\sum_{n\ge1}(D/H)^n/(n\,n!)=\tfrac12\E(D/H)\). This is the finite angular term produced by the subtraction.

For the nuclear-weighted phase put $T_\beta(q)=\widetilde{(\ee^{-d_p b^2}t)}(q)$ and define
\begin{align}
 c_\beta&=\frac{\beta_p^4}{a_0^2+2\beta_p^2},&
 M&=\frac{\beta_p^4}{b_0^2},&
 C_b&=a_0^{-2}+d_p. \label{eq:cdefs}
\end{align}
\begin{align}
 T_\beta(q)&=2\pi\beta_p^2\ee^{-g_p q^2}
    [c_C+\ln(M/c_\beta)-\gamma_E-\E(c_\beta q^2)]
  \notag\\
  &\quad+\frac{2\pi\kappa}{C_b}\ee^{-q^2/(4C_b)}.
 \label{eq:tbeta}
\end{align}
With $A=a+g_p$, $A_1=a+1/(4C_b)$ and $z_\beta=Dc_\beta/[A(A-c_\beta)]$, the convolution of Eq.~\eqref{eq:tbeta} becomes
\begin{align}
 \mathcal M_{t\ee^{-d_p b^2}}(a,L)
 &=\frac{2\pi^2\beta_p^2}{A}\ee^{D/A}
       \left[c_C-\gamma_E-\E(z_\beta)\right.\notag\\
 &\qquad\left.+\ln\frac{M(A-c_\beta)}{Ac_\beta}\right]
       +\frac{2\pi^2\kappa}{C_b A_1}\ee^{D/A_1}.
 \label{eq:mtbeta}
\end{align}
Indeed, letting $A_u=A-u c_\beta$, the Gaussian integral of $\E(c_\beta\ell^2)$ divided by $\pi\ee^{D/A}/A$ is
\begin{align}
 &\int_0^1\frac{\mathrm{d}u}{u}
   \left[\frac{A}{A_u}\ee^{D/A_u-D/A}-1\right]\notag\\
 &\qquad=\ln\frac{A}{A-c_\beta}+\E(z_\beta).
 \label{eq:u}
\end{align}
In the part proportional to \((\exp[\cdot]-1)/u\), the substitution \(v=u/(A-u c_\beta)\) gives the integral for \(\E\). The remaining term gives a logarithm. Since \(a>0\) and \(c_\beta/g_p=2\beta_p^2/(a_0^2+2\beta_p^2)<1\), one has \(A>c_\beta\) without an assumption of a small profile width. With \(U_\eta\to\ii\eta t(1-\Gamma_p)\), Eqs.~\eqref{eq:mt}, \eqref{eq:tbeta}, and \eqref{eq:mtbeta} give the first-order corrections in Eq.~\eqref{eq:feta}. The coordinate integral~\eqref{eq:moment} gives the same moments when the phase constant, including the finite part of Eq.~\eqref{eq:screen}, is kept.
\subsection{Screened normalization and finite charge}
For a screened point charge the potential is proportional to \(\ee^{-\mu r}/r\). Along the straight trajectory of impact parameter \(b\), put \(z=b\sinh u\); the dimensionless phase is \(-2\int_0^\infty\ee^{-\mu b\cosh u}\dd u=-2K_0(\mu b)\), with \(K_0\) the modified Bessel function of the second kind. For small \(\mu b\), \(-2K_0(\mu b)=2\ln(\mu b/2)+2\gamma_E+o(1)\). Adding \(d_\mu\) of Eq.~\eqref{eq:screen} gives \(t_{\ptc}\) as \(\mu\to0^+\). The added constant is the zero mode in momentum space, so its finite part must be fixed in the same phase convention as the other terms.

Multiplication by $F_{\rm ch}(Q)$ adds a nonsingular transform to the point result. The part proportional to $1-\ee^{-hQ^2}$ is seen directly from
\begin{equation}
 \frac{1-\ee^{-hQ^2}}{Q^2}=\int_0^h\ee^{-sQ^2}\dd s ,
 \label{eq:chargesplit}
\end{equation}
whose inverse transform, multiplied by $4\pi$, is $\int_0^h \ee^{-b^2/(4s)}\dd s/s=E_1(b^2/(4h))$. The remaining Gaussian Fourier term, $2\pi\kappa a_0^2\ee^{-hQ^2}$, transforms to $2\kappa\ee^{-b^2/a_0^2}$. This derives Eq.~\eqref{eq:phase} with the same reference phase as Eq.~\eqref{eq:screen}.

\subsection{Relation to a Laguerre derivative}
The angular subtraction also has a representation as a derivative with respect to the degree of the Laguerre function. For real \(B\), \(H>0\), and \(\mathrm{Re}\,s>0\), the series for \(J_0\) gives
\begin{align}
 &\int_0^\infty \ell^{2s-1}\ee^{-H\ell^2}
                  [J_0(B\ell)-1]\dd\ell \notag\\
 &\quad=\frac{\Gamma(s)}{2H^s}
       \big[{}_1F_1(s;1;-B^2/(4H))-1\big].
 \label{eq:lag}
\end{align}
Here \({}_1F_1\) is the confluent hypergeometric function. The subtracted integral exists at \(s=0\). Since \(\Gamma(s)\sim1/s\) and \({}_1F_1(0;1;z)=1\), its limit is \(\tfrac12\,\partial_s{}_1F_1(s;1;z)|_{s=0}=\tfrac12\E(z)\), with \(z=-B^2/(4H)\). In terms of \(L_\nu(z)={}_1F_1(-\nu;1;z)\) the limit is \(-\tfrac12\,\partial_\nu L_\nu(z)|_{\nu=0}\). Using \(I_0\) instead of \(J_0\) changes the sign of \(z\). The derivative gives the subtracted integral; the part removed from the integrand is accounted for by Eq.~\eqref{eq:screen}.
\section{Closure for the two breakup channels}\label{app:closure}
At fixed \(\vct B\), the operator \(\widehat{\mathcal S}_\eta\) multiplies the internal state by \(S_pS_n\ee^{\ii\eta t}\). Denote the expectation in \(\psi_S\) by \(\langle\cdots\rangle_S\). Integrating the projected plane waves with measure \(\dd^3k/(2\pi)^3\) gives \(1-|\psi_S\rangle\langle\psi_S|\). In the frozen approximation the coherent breakup probability, summed over the final internal momenta, is therefore
\begin{equation}
 P_{\coh}(\vct B)=\langle|\widehat{\mathcal S}_\eta|^2\rangle_S
               -|\langle\widehat{\mathcal S}_\eta\rangle_S|^2 .
 \label{eq:closure}
\end{equation}
The first expectation value is independent of the Coulomb phase, but the elastic projection in the second depends on it. The identity part of the operator cancels from the variance. Writing \(P_p=1-|S_p|^2\) and \(P_n=1-|S_n|^2\) at the nucleon impact parameters, we have \(1-|S_pS_n|^2=P_n|S_p|^2+P_p|S_n|^2+P_pP_n\). Only the first term represents neutron removal with the proton surviving.

In particular, restoring the factor $\ee^{-\ii\vct p_\perp\cdot\vct r_\perp}$ in the spectator amplitude and integrating the momentum probability over the entire proton momentum space gives, by Parseval's identity,
\begin{align}
 \sigma_{\remn}
  &=\int\dd^2b_n\,P_n(b_n)\notag\\
  &\quad\times\int\dd^3r\,
           |\psi_S(r)|^2\,|S_p(|\vct b_n+\vct r_\perp|)|^2 .
 \label{eq:pars}
\end{align}
The integrated removal probability is unchanged by the phase of unit modulus. The proton yield at a given momentum is a Fourier intensity, so the phase can change its shape, as in Table~\ref{tab:changes}. The closure identities~\eqref{eq:closure} and \eqref{eq:pars} do not imply cancellation between the two Coulomb changes in any one bin.
\section{Kernels at finite longitudinal and transverse momenta}\label{app:finite}
\subsection{Coherent terms containing the nuclear profiles}
Let $T_D=\widetilde D_\eta$, $T_{pD}=\widetilde{(\ee^{-d_p b^2}D_\eta)}$, and
\begin{align}
 \Phi_\zeta(\vct\ell)&=\ee^{-g_n|\vct Q-\vct\ell|^2}
                   G(\vct q_{\ell,\zeta},\vct k),\notag\\
 \vct q_{\ell,\zeta}&=(\vct Q/2-\vct\ell,\zeta/2).
 \label{eq:phiz}
\end{align}
The label $\zeta$ takes the values $0$ and $Q_z$ and specifies the longitudinal argument of the transition form factor. The strong amplitude is obtained from Eq.~\eqref{eq:fn} by $C_N\to\overline C_N$, $G_\pm\to G_\pm^z$, and $\Phi\to\Phi_0$. The terms with at least one nuclear profile in the Coulomb correction are
\begin{align}
 \Delta\F_{\eta}^{\rm nuc}
 &=\frac{\ii\overline C_p}{2\pi}T_{pD}(Q)G_-^z
 \notag\\
 &\quad+\frac{\ii\overline C_n\beta_n^2}{4\pi^2}
        \int\dd^2\ell\,T_D(\ell)\Phi_{Q_z}(\vct\ell)\notag\\
 &\quad-\frac{\ii\overline C_p\overline C_n\beta_n^2}{4\pi^2}
        \int\dd^2\ell\,T_{pD}(\ell)\Phi_0(\vct\ell).
 \label{eq:shortz}
\end{align}
The second line contains the longitudinal neutron phase; in the last line the opposite proton and neutron phases cancel. The internal \(\ell\) convolution uses the same distributional subtraction as at \(Q_z=0\). The \(|Q_z|\) scale in Eq.~\eqref{eq:joined} enters the external pole only.

Using Eq.~\eqref{eq:wn}, each plane-wave Gaussian in $\Phi_\zeta$ is
\begin{align}
 &w_j\,\ee^{-\nu_j(k_z+\zeta/2)^2-g_n Q^2-\nu_j p_\perp^2}
 \notag\\[-2pt]
 &\qquad\times
 \exp[-(g_n+\nu_j)\ell^2
       +(2g_n\vct Q+2\nu_j\vct p_\perp)\cdot\vct\ell] .
 \label{eq:finitegauss}
\end{align}
The density term has the width and vector of Appendix~\ref{app:gauss} and acquires the factor \(\ee^{-\xi_{ij}\zeta^2/4}\) from the longitudinal argument of \(n_S\). Equation~\eqref{eq:moment} applies with the full norm of the vector in Eq.~\eqref{eq:finitegauss}. The angle \(\vartheta\) between \(\vct Q\) and \(\vct p_\perp\) enters through \(\vct Q\cdot\vct p_\perp\). We square the amplitude at each \(\vartheta\) and then integrate over that angle; averaging \(\vct k_\perp\) before calculating the amplitude changes the spectrum.

For real optical profiles (\(\rho_N=0\)), simultaneous inversion of \(\vct B\) and \(\vct r\) conjugates the longitudinal factors of Eq.~\eqref{eq:profilesz}. Since \(G_T\) and \(G_L\) are real, \(\F_{-\eta}=-\F_\eta^*\) also holds with Eq.~\eqref{eq:joined}. The coherent spectrum is therefore even in \(\eta\) at nonzero angle within this model.
\subsection{Azimuthal reduction of the spectator integral}
For the $j$th Gaussian, write $\vct b'=\vct b_n+\vct r_\perp$, $b=|\vct b_n|$ and $\varphi=\angle(\vct b_n,\vct p_\perp)$. Apart from the overall factor $\ee^{\ii\vct p_\perp\cdot\vct b_n}$, which cancels in the spectrum, the angular integral over $\vct b'$ is
\begin{align}
 J_j(b',\varphi)&=2\pi I_0(b'B_j),\notag\\
 B_j^2&=(2\lambda_j b)^2-p_\perp^2
           -4\ii\lambda_j b p_\perp\cos\varphi .
 \label{eq:complexI}
\end{align}
Integrating the exponential \(\exp[b'\hat{\vct b}'\cdot(2\lambda_j\vct b_n-\ii\vct p_\perp)]\) over its azimuth gives Eq.~\eqref{eq:complexI}. Either branch of the square root may be used because \(I_0\) is even. The two radii are integrated numerically, and the squared amplitude is averaged over \(\varphi\). At \(p_\perp=0\) Eq.~\eqref{eq:complexI} reduces to the angular factor in Eq.~\eqref{eq:radrem}. At \(\eta=0\) the radial Gaussian integral can also be evaluated at nonzero \(p_\perp\):
\begin{align}
 R_j^{[0]}(b,\varphi)
 &=\frac{\pi}{\lambda_j}\ee^{-p_\perp^2/(4\lambda_j)-\ii u_b}
     \notag\\
 &\quad-\frac{\pi C_p}{\lambda_j+d_p}\ee^{-X_j},\notag\\
 X_j&=\frac{\lambda_j d_p b^2+
                   p_\perp^2/4+\ii\lambda_j u_b}{\lambda_j+d_p},\notag\\
 u_b&=b p_\perp\cos\varphi .
 \label{eq:remcheckfinite}
\end{align}
The same overall phase has been removed from both sides. In the full integral we combine the real exponential from \(I_0\) with the radial Gaussian to keep intermediate values bounded at large radii. The analytic result~\eqref{eq:remcheckfinite} checks the complex amplitude before it is squared.
\section{Coulomb contributions presented at the Kyiv conference}\label{app:conf}
\begin{figure*}[t]
 \centering
 \includegraphics[width=\textwidth]{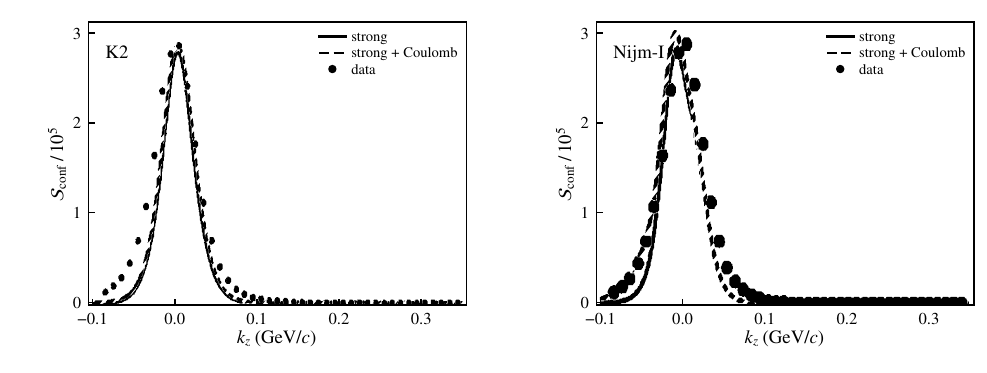}
 \caption{Conference calculations of the Coulomb contribution to deuteron breakup on carbon \cite{BITP2026}: K2 (left) and Nijm-I (right). The solid curves are strong calculations and the dashed curves include the Coulomb contribution. The markers and ordinate notation are those of the conference illustrations.}
 \label{fig:conf}
\end{figure*}

The two panels in Fig.~\ref{fig:conf} were presented at the Bogolyubov Kyiv Conference \cite{BITP2026}. They show how the Coulomb--nuclear terms alter the calculated breakup spectra on carbon for the K2 and Nijm-I inputs used in that presentation. The ordinates \(\mathcal S_{\rm conf}\) and the markers follow the conventions of the conference figures. The zero-angle comparison in the main text uses the common inputs specified there.

\begingroup
\fontencoding{T2A}\fontfamily{cmr}\selectfont
\raggedright\rezume{\ifdefined\ReviewCopy\else Я.~Д. Кривенко-Еметов, Б.~І. Сидоренко\fi}
{Розвал дейтрона на вуглеці та водні: кулонівські ефекти й надлишок при великих імпульсах}
{\AbsUA}
{Ключові слова: розвал дейтрона, спектр швидких протонів, кулонівська фаза, модель Глаубера--Ситенка, видалення нейтрона.}
\endgroup
\end{document}